\documentclass[a4paper,alpha-refs]{aejstyles}

\journal{aej}

\usepackage{graphicx}
\usepackage{siunitx}
\newcommand{\cmmnt}[1]{}
\usepackage[left]{lineno}
\title{Teaching the apparent motion of Sun and stars across four European countries}

\author[1]{Hans BEKAERT}
\author[2]{Wim VAN DOOREN}
\author[3]{Hans VAN WINCKEL}
\author[4]{Markus POESSEL}
\author[5]{Inge THIERING}
\author[6]{Marco NICOLINI}
\author[7]{Enrico ARTIOLI}
\author[8]{Despina AVGERINOU}
\author[9]{Eleana BALLA}
\author[1,\authfn{1}]{Mieke DE COCK}

\affil[1]{KU Leuven, Department of Physics and Astronomy and LESEC, Celestijnenlaan 200c – box 2406, 3001 Leuven, Belgium}
\affil[2]{KU Leuven, Centre for Instructional Psychology and Technology, Dekenstraat 2 – box 3773, 3000 Leuven, Belgium}
\affil[3]{KU Leuven, Department of Physics and Astronomy, Celestijnenlaan 200d – box 2401, 3001 Leuven, Belgium}
\affil[4]{KU Leuven, Department of Earth and Environmental Sciences and LESEC, Celestijnenlaan 200e – box 2409, 3001 Leuven, Belgium}
\affil[4]{Haus der Astronomie, Königstuhl 17, D-69117 Heidelberg, Germany}
\affil[5]{Max-Born-Gymnasium Neckargemuend, Alter Postweg 10, 69151 Heidelberg, Germany}
\affil[6]{Liceo Scientifico Statale "A. Tassoni", Viale Virginia Reiter 66, 41121 Modena, Italy}
\affil[7]{Civico Planetario F. Martino di Modena, Viale Jacopo Barozzi 31, 41124 Modena, Italy}
\affil[8]{1st Experimental Junior High School of Thessaloniki, Ethnikis Aminis 26, 546 21 Thessaloniki,  Greece}
\affil[9]{Noesis - Thessaloniki Science Center and Technology Museum, 6th km Thessaloniki-Thermi Rd, 57001 Thermi, Greece}

\papercat{Research Article}

\runningauthor{First et al.}

\jvolume{00}
\jnumber{0}
\jyear{2024}

\begin{document}

\begin{frontmatter}
\maketitle
\begin{abstract}
In the context of the European Erasmus+ project Teaching ASTronomy at the Educational level (TASTE), we investigated the extent to which a learning module at school and a set of activities during a planetarium visit help students to gain insight in the Apparent Motion of the Sun and Stars. Therefore, we have set up a two treatment study with a pretest posttest design. In the four participating countries (Belgium, Germany, Greece and Italy), secondary school students studied the concept of the celestial globe at school using newly designed learning materials. The first group with the Belgian students only took this learning module at school. A second group with the German, Greek and Italian students applied the concepts learned at school during two hands-on workshops and a planetarium presentation in the dome. Before the start of the learning activities at school, we administered the Apparent Motion of Sun and Stars (AMoSS) test as pretest with twelve multiple choice questions to the 13-17 years old students of eight European secondary schools (\textit{N}=262). Immediately after the learning activities in school or planetarium, the students took the AmoSS test again (posttest). On average, the results of both tests show similar learning gains for the two groups, but we see a clear distinction between the Sun-related and the star-related questions. In the first group, the learning gains on questions about the stars are higher than on questions about the Sun, while in the second group it is just the other way around. We also see differences between the two groups, when looking at the learning gains per question. By using a latent class analysis, we identified different classes of student answers. We show how students evolve from one class to another between pretest and posttest. Overall the results of the pretest and posttest show that a good understanding of the different aspects of the apparent motion of celestial bodies is difficult to achieve.
\end{abstract}


\end{frontmatter}

%%% Key points
%\begin{keypoints}
%\begin{itemize}
%\item This is the second point
%\item One last point.
%\end{itemize}
%\end{keypoints}

\section{Introduction}
\label{sec:introduction}
A lot of schools visit planetariums to observe celestial motions, learn the constellations or study the phases of the Moon.  Therefore, it's worthwhile to investigate if a planetarium is a a good environment for teaching astronomy. Research on this topic suggests that a planetarium not necessarily is a better environment than the classroom \cmmnt{(REF article (414) Brazell & Espinoza 2009)}(\cite{2009_Brazell}): some studies conclude that the classroom has significant advantages over a planetarium \cmmnt{(REF article 416 p. 46 Reed 1670a, Smith 1966 (see ART (1) on page 55)}(\cite{1966_Smith,1970_Reed}), while other studies show that the planetarium offers opportunities that outperform the classroom \cmmnt{(REF article 416 p. 46 Plummer, Yu)}(\cite{2009_Plummer,2014_Plummer,2015_Yu}).
The European Erasmus+ project Teaching ASTronomy at Educational level (TASTE), which is a collaboration between four planetariums or science centres, a university and four schools, aims at identifying essential elements for teaching basic astronomical concepts both in the classroom and the planetarium. Since the project runs in four European countries (Belgium, Germany, Greece and Italy) we are especially interested in features of the learning environment that are ‘universally’ important, independent of the local educational system.  As the main topics are of this Erasmus+ project are time and space, we decided to focus on the apparent motion of the Sun and stars.

In the first phase of the project, we tested students' knowledge and understanding of this topic in the four participating countries. We have reported on this first study in a previous manuscript (\cite{2022_Bekaert2}: we found similar results in all countries. Despite the fact that all participating students have been taught some basic astronomical concepts in school, they only demonstrated a rudimentary understanding of the apparent motion of the Sun and stars during the day, during the year and for different locations of the observer on Earth. Moreover, we saw a clear distinction between the responses to the Sun-related and the star-related questions. In general, the questions about the Sun were answered more correctly than the questions about the stars. Based on these findings we concluded that specific instructions are needed to improve students’ understanding of the apparent celestial motions.

In the second phase of the project, teachers and planetarium staff collaboratively designed a learning intervention that includes new teaching/learning materials to support student understanding of crucial elements like spatial scales and time, related to the apparent motion of the Sun and stars. This resulted in a learning module where the concept of the celestial sphere is introduced, two planetarium workshops where students learn to use 2D- and 3D-models of Earth and the celestial sphere, and a planetarium presentation in the dome. We tested how these learning activities helped the students to gain insight in the apparent motion of celestial bodies.

In this article, we report our findings on how the newly designed planetarium workshops, planetarium presentation and learning module at school support student thinking about the apparent motion of the Sun and stars, based on the administration of the Apparent Motion of the Sun and Stars (AMoSS) test (\cite{2020_Bekaert,2022_Bekaert2}) to a group of 13-17 years old students of four European countries (Belgium, Germany, Greece and Italy). 

In Section 2, we discuss earlier research on students' difficulties in learning the apparent motion of the Sun and
stars and on the use of a planetarium in education. In Section 3, we formulate the research questions and in Section 4 we describe the design of the intervention, how we tested its impact and how we analyzed the answers of the multiple choice questions. Section 5 lists the results of the various steps in the study. The last sections conclude with a discussion and implications for education.

\section{Background}
\label{sec:background}

\subsection{Students' difficulties in learning about the apparent motion of the Sun and stars}

Although most people are fascinated by astronomy, in the literature many studies indicate that young children, students and  adults have difficulties in developing a deep understanding of basic astronomical concepts \cmmnt{(zie ART 1 REF 2-7)}(\cite{2011_Plummer,2003_Bailey,2009_Lelliott,2015_Testa,2000_Trumper,2020_Bekaert}). Students of different ages and coming from different places in the world often show alternative ideas to explain the apparent motion of the Sun and stars. In previous manuscripts (\cite{2022_Bekaert,2022_Bekaert2}), we described different alternative ways of student thinking: 
\begin{enumerate}
		\item Many students think that stars behave exactly like the Sun: for both the Sun and the stars the path of the apparent motion is higher and wider in summer than in winter. When the observer’s latitude decreases, this path becomes higher and wider both for the Sun and the stars.
        \item Students sometimes think that star trails behave opposite to the Sun’s path. They argue that in winter the star trails are higher and wider, because the nights are longer. Also, when the observer’s latitude increases, the star trail becomes higher and wider. 
		\item Concerning the annual motion students often think that the Sun’s path does not change throughout the year. They have problems to understand how the Sun's culmination height and the position of sunrise and sunset depends on the time of the year and the observer's latitude.
		\item Many students are not aware of the fact that the star trails stay fixed during the year and they do not know how star trails change if the observer changes position.
		\item Students have difficulties with the correct interpretation of the culmination height of the Sun and the stars: they think that that the culmination height is proportional to the observer's latitude. 
\end{enumerate}

These various alternative ideas may be an expression of the mental models students use when answering questions about the apparent motion of the Sun and stars. There is some debate in the literature about the exact definition of a mental model, but in accordance with constructivism, a mental model refers to an internal representation of the external world, containing meaningful declarative and procedural knowledge that people use to understand specific phenomena \cmmnt{(zie studietekst constructivsime}(\cite{2012_Aldiban}). In our research, we use the definition of \cite{2011_Corpuz} as a guideline. They define a mental model as a “students’ way of understanding a certain physical phenomenon,” which can also be an unseen physical phenomenon. We follow \cite{1998_Gilbert} who argue that a mental model as such is inaccessible and that as researchers we can only rely on an expressed version of it. Mental models are not permanent and may (gradually) change over time, and their expression is formed on the spot, for example in a written explanation or a drawing, in response to a given prompt or question. \cmmnt{(REF 8 in Art. 2)}According to \cite{1987_Collins_BOOK}, mental models can be formed through analogical thinking\cmmnt{(see Art. 3 or Art. 2)}: when describing a concept with which a student is unfamiliar, he/she tends to make a comparison with an equivalent concept with which he/she is more familiar and is perceived as similar.

Despite lessons at school or in the planetarium about the Earth's rotation, the Earth's revolution and the Earth's shape, students struggle with linking these concepts to the apparent motion of the Sun and stars (\cite{2022_Bekaert2}). This seems to be due to the fact that connecting the observed celestial motions from a geocentric point of view to the allocentric view from space\cmmnt{(zie art (377) p 2. Nussbaum} is not an easy process. In the literature, it is suggested that it seems to be essential that students learn to think and alter between a geocentric and an allocentric frame of reference in order to understand the apparent motion of the Sun and stars and link these to the actual motion of the Earth \cmmnt{(zie (ART 1) REF [2][3][22][28])}(\cite{2011_Lopresto,2010_Plummer,2015_Testa,2015_Yu}). Probably, specific instructional strategies are needed: students must be trained to switch between different frames of reference, to be able to really understand apparent celestial motions. In our research project, we investigate how a school trip to a planetarium can support this process.

\subsection{Teaching astronomy in the planetarium}
Since the first planetarium began operations at the Deutsches Museum in Munich in 1925, planetariums have been used for teaching the fundamentals of astronomy. As for the effectiveness of the planetarium, systematic studies have painted a mixed picture: In his doctoral dissertation \cmmnt{(zie (1) Slater p 7}\cite{2009_Brazell}  explored the instructional effectiveness of the planetarium in astronomy education through a meta-analysis of 19 studies. He found that the planetarium has not been a very effective tool for improving student attitudes towards astronomy. In their review study on teaching astronomy in the planetarium  \cite{2017_Slater_BOOK}\cmmnt{(zie (1) Slater p 7} nuance this conclusion and state that "the planetarium has been statistically effective” (Slater and Tatge, 2017, p. 7). We list some interesting studies that confirm this nuance and are relevant to our research.

Several studies compare the performance of students taking well designed astronomy classes at school with students taking a trip to the (traditional) planetarium. For example, \cmmnt{(zie (1) Slater p 61} \cite{1970_Reed} specifically measured students’ understanding of daily and yearly motions of the Sun and stars using an expert validated test. He measured the effectiveness of the planetarium as a teaching facility compared to a classroom instructional setting where he used student-manipulated celestial spheres during instruction. He also administered a retention test 8 weeks after instruction to determine the durability and retention of students’ understanding. His data on 159 college students suggested that there was no statistically significant difference in student scores in the two treatments. He concluded that for teaching, the planetarium dome itself is best used not as a passive theater where seats are reclined and the lights are low, but as a teaching classroom, using all the best teaching practices and pedagogical techniques available.   

After Reed’s findings that the planetarium instructional interventions needed to adopt the best practices for effective science teaching,\cmmnt{(zie (1) Slater p 70} \cite{1982_Edoff} conducted a two-group comparison study of 542 5th and 8th grade students to determine if using manipulatives under the planetarium dome would enhance students’ achievement. In his study, both groups of students learned astronomy in the planetarium, but only one group did so by individually manipulating learning materials as for example a celestial sphere. He studied three big ideas in astronomy: (i) celestial sphere and time ; (ii) seasonal changes; and (iii) lunar movement and moon phases. He found that the proper use of manipulatives resulted in greater learning gains as opposed to students who only attended the planetarium session. 

In an effort to increase learning gains, \cmmnt{(zie (1) Slater p 78} \cite{2000_Meyer} investigated the value of pre- and post-planetarium visit learning activities. She designed a multiple treatment, pretest posttest study to see if students who participated in various pre- and post-visit activities would increase the number of items they answered correctly on their test. One treatment group experienced hands-on activities before and after their planetarium visit, a second treatment group used audio-visual activities before and after their visit, a third treatment group completed text and reading based activities before and after the planetarium. The learning gains were small (about 15 \%) and there were no significant differences among the various treatment groups. 

Rusk's attempt to make a planetarium visit more effective by adding a post-visit activity to it, was more successful \cmmnt{(zie (1) Slater p 81} (\cite{2003_Rusk}). Using a large scale, two-group, comparison study, he first showed students a pre-recorded planetarium program about phases of the moon (Moonwitch, written by Phil Groce for Bowen Productions). Following the show, half of the students completed a hands-on activity using light bulbs and styrofoam balls before taking the posttest while the other half of students took the posttest without attending the post-visit activity. Students who completed the activity scored better than students who had not. He interpreted this as evidence of the critical importance of using hands-on activities as part of instruction.

In their multitreatment study with 3rd grade students \cmmnt{(zie (1) Slater p 96 - REF (38) p. 21 } \cite{2013_Plummer} draw the same conclusion. Their study clearly showed that elementary students have the greatest learning
gains when they engage in instruction that supports their ability to visualize Earth-based observations: observe simulations, guided gesturing and participate in kinesthetic modeling. They concluded that whether or not the planetarium is part of instructions is not important. Instead, it is how the planetarium is used best as learning environment: showing celestial motion to actively engaged students. The digital planetarium can play an interesting role in supporting students, especially because it can show both the geocentric and allocentric (the view from space) viewpoints.

To summarize, the above studies show, as \cite{2017_Slater_BOOK} conclude in their review study on teaching astronomy in a planetarium, \cmmnt{(zie (1) Slater p 123} that "the planetarium is unarguably able to capture attendees’ innate interest, but that lasting change requires purposeful educational decisions in order to be relevant and effective". 
The aim of this study is to investigate the effect of a specific, purposeful designed intervention on students' understanding of the apparent motion of the Sun and stars.

\section{\label{sec:level2}Research questions}
In order to support students’ learning on the apparent motion of the Sun and stars, we designed several teaching/learning activities, both to implement at school (which we call ‘learning module’) and in the planetarium (which we call ‘planetarium activities’). The main research questions we answer in this study are
\begin{enumerate}
		\item To what extent does the TASTE learning module at school lead to learning gains among secondary school students regarding the apparent motion of the Sun and stars?
              \item To what extent does a sequence of the TASTE learning module at school, followed by planetarium activities, lead to learning gains among secondary school students regarding the apparent motion of the Sun and stars?
	\end{enumerate}
 
To answer these questions, we used a two-treatment study with pretest posttest design. The AMoSS test (\cite{2020_Bekaert, 2022_Bekaert2}) was used as test instrument.

\section{Methods}

\subsection{\label{sec:level2}Participants}
Two groups of students participated in the study. The first group consisted only of Belgian students ($N = 90$). These came from three different public schools, but were all in the 11th grade of secondary education (16-17 year olds). Their curriculum consists of several science courses (physics, biology, chemistry geography) and 6-8 hours mathematics a week. Before participating in the study, these students had learned in geography classes about the Earth's rotation around its axis, the Earth's revolution around the Sun and about the apparent motion of the Sun.

The second group consisted of students from Germany, Greece and Italy. The German students ($N=30$) were in the 8th and 9th grade of secondary education (13-15 year olds) of one public school. They were all taking a science course with 4 hours of mathematics per week. Before participating in the study, these students had learned about the Moon phases and lunar and solar eclipses. 

The Greek students ($N=46$) were in the 8th grade of secondary school (13-14 year olds) of one model state junior high school. They take an exam in math and language after primary school in order to be accepted. The students all took a course with 4 hours of mathematics. Notions about the Earth's revolution around the Sun, the Earth's axis tilt and seasons are only picked up in 1st grade's geography book in one paragraph which is not compulsory. 

The Italian students ($N = 96$) came from three different schools, one of which was a private school. This private school is officially recognized by the Italian ministry of education, as a school with the same public syllabus in any subject and following the same public rules about students, teachers and procedures, with a final state exam. Most students of this private school were in the 9th grade of secondary education (14-15 year olds), a few of them were in grade 11 (16-17 year olds). They all took a science course with 5 hours of mathematics. The students of the second Italian school were all in the 9th grade and were all taking a science course with 5 or 6 hours of mathematics. The students of the third Italian school took  different courses with 2 to 5 hours of mathematics, in the 9th grade of secondary education. Only these students had also learned in class about the Earth's rotation around its axis, the Earth's revolution around the Sun and about the apparent motion of the Sun.

\begin{table*}[]
\caption{\label{tab:table3}Participating students of group 1 (\textit{N}=90) and group 2 (\textit{N}=172)}
\centering
%\begin{ruledtabular}
\renewcommand{\arraystretch}{1.3}
\begin{tabular}{p{1.8cm} p{1.8cm} c c c c c c} \hline
Group & Country & Number 		&Number       		& Number  			&Male 	&Female   & Age \\
      &         & of schools 	& of class groups 	& of students  		&		&		&\\ \hline
Group 1 &Belgium   &3 &6  &90 &37 &53    &15-16
    \\ \hline
Group 2 &Germany  &1 &2  &30  &17 &13   &13-14-15
    \\
&Greece   &1 &2 &46 &21 &25    &13-14
    \\
&Italy  &3 &6  &96 &58 &38    &14-15-16-17
    \\ \hline
\end{tabular}
%%\end{ruledtabular}
\end{table*}

Table 1 lists all the details of the participating students. They were free to decide whether or not to participate. Only the students who signed the informed consent form (or had their parents sign the form, depending on local legislation) and who participated in both the pretest and the posttest are included in the study. No incentive was given to the students. 

\subsection{\label{sec:level2}Design of learning and teaching materials}
Based on the available research literature, we listed the following underlying design principles, which essentially are related to a constructivist learning perspective (\cite{2002_Burns}).
\begin{enumerate}
\item We explain the similarities and differences between the Sun and the stars systematically (\cite{2020_Bekaert,2022_Bekaert}).
\item We encourage students to think in and switch between different reference systems: the geocentric point of view and the allocentric point of view (\cite{2014_Plummer}).
\item We use a physical 3D model of the celestial sphere, whose different parts (observer’s horizon, equator, ecliptic, ...) are introduced and explained step by step (\cite{1973_Reed,1982_Edoff}).
\item We minimize teacher instructional time and encourage students to reflect and discuss in order to make their existing mental models explicit to each other while making exercises, and to stimulate model-based thinking.
\end{enumerate}

We focused on the fact that students performed better on the Sun-related questions than on the star-related questions and that students usually do not use a proper scientific model in explaining their choices. As a remedy, we decided to teach the students about the celestial motions using the model of the celestial sphere. Since we believe that a planetarium is an ideal tool to visualize the different aspects of this 3D model, we developed two planetarium workshops and a new planetarium presentation. The workshops were organized by the planetarium staff at the planetarium outside the dome. To prepare for these planetarium activities, we created a learning module that students worked through at school before visiting the planetarium.
Teachers and planetarium staff worked together during the design process. At the start of this process, they defined the following constraints: 
\begin{enumerate}
		\item The activities should not assume any specific prior knowledge of the phenomena that are studied in the module. They must be accessible to secondary school students of 13-18 years old.
            \item Because science teachers are constantly under time pressure, the learning module at school should be completed in two standard lessons of 40 minutes each.
		\item All activities should be hands-on, so that students actively engage with the concepts being taught.
            \item To schedule the various planetarium activities on one school day, the workshops and planetarium presentation should not exceed 4 hours.
            \item The planetarium presentation should be suitable for display in both analogue and digital planetariums.
            \item All activities should be presented in the students' native language.
	\end{enumerate}
 The development of the learning and teaching materials was an iterative process. Two test runs were implemented in which students and teachers from the different participating countries were asked to provide feedback. Based on this feedback, the materials were adapted and fine-tuned. A protocol was provided for the teachers and planetarium staff so that the activities were run in the same way across countries.

 All learning materials and the scenario of the planetarium presentation are freely available on line (see Appendix A).

\subsubsection{Learning module}
The learning module consists of two parts: (1) the daily motion of the Sun and the stars, and (2) the Sun's path and the star trails throughout the year. 

In the first part, students learn to explain the daily motion of the Sun and stars. Through exercises, they learn that the apparent motion of celestial bodies is due to Earth's rotation around its axis. The model of the celestial sphere is introduced step by step. Concepts as the observer's local horizon, cardinal directions, Sun's path and star trail, the relation between the observer's latitude and the altitude of Polaris... are practised.

The second part is about the Earth's revolution and how this yearly motion is related to the change of the Sun's culmination height and the position of rise and set throughout the year. Students learn why this is not the case for the stars. The concept of the ecliptic and the ecliptic plane is introduced and students learn that while the stars have fixed positions on the celestial sphere, the Sun moves apparently on the ecliptic. In the exercises, students discover why certain constellations are not visible every night. For an example, see Figure 1.
The exercises pay special attention to perspective-taking: in some exercises they take the geocentric point of view, and in other exercises they take the allocentric point of view.

\begin{figure}[h] % de h staat voor "hier"
\centering
\includegraphics[width=8.5cm]{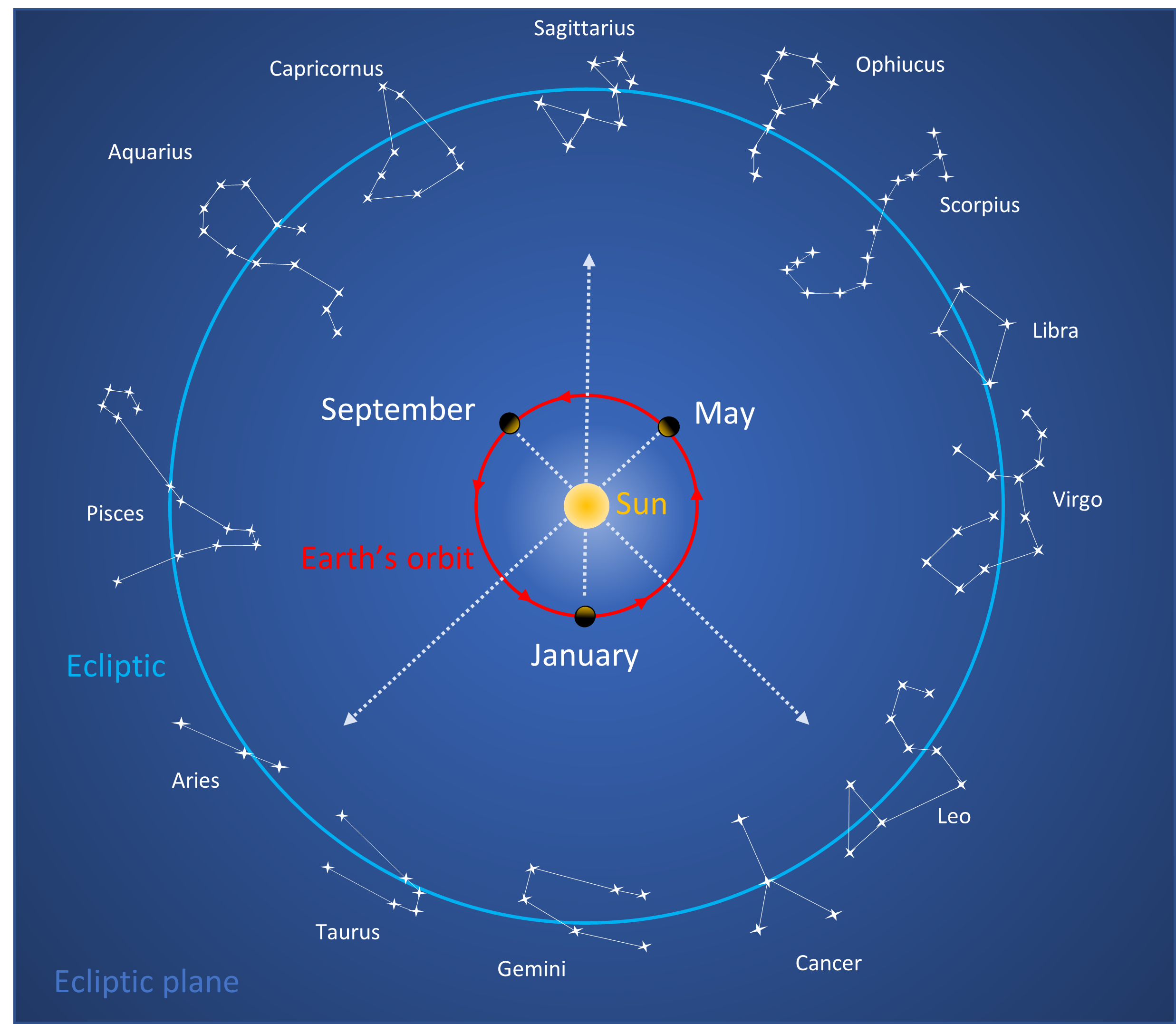}% Here is how to import EPS art
\caption{\label{fig:epsart} Example of an exercise: Can you observe the constellation Leo during September? Why (not)?}
\end{figure}

\subsubsection{Planetarium workshop 1: 3D model of the Earth and the celestial sphere}
During the first planetarium workshop the students practice the concepts, which they learned during the preparatory learning module at school, by using 3D models of the Earth and the celestial sphere. They focus on an allocentric point of view.

In the first part the concept of the parallel globe is introduced. The parallel globe is a 3D model of the Earth, on which you can place different observers using magnets. For each observer’s location the globe can be oriented in such a way that the horizontal plane under the observer's feet is parallel to the tangent plane to the globe in its upper point. In this way for each location, the direction of the globe's axis is parallel to the Earth’s axis and the poles of the globe point toward the celestial poles. Using the parallel globe the students can find their place on Earth, they locate the local horizon and meridian, as well as the cardinal points at their exact location using the polar star. On a sunny day, you can compare the shadows of observers at different locations on Earth. Through exercises, students discover that Polaris' altitude varies from one location to another and that it indicates the observer's latitude.

In the second part students use a 3D model of the celestial sphere. The model is called "bottle globe" (\cite{2016_Fisher}) and was built with easy to find materials: a round flask (celestial sphere), a stick (Earth's axis), a ping-pong ball (Earth) and water (horizon). The celestial equator and the ecliptic are indicated on this model. By changing the inclination of Earth's axis, students practice to locate Polaris for different locations of the observer (See Figure 2). Through brief exercises, students locate the Sun on the ecliptic for different times throughout the year and they discover that the Sun's path depends on this position, while the star trails do not change throughout the year.

\begin{figure}[h] % de h staat voor "hier"
\centering
\includegraphics[width=8.5cm]{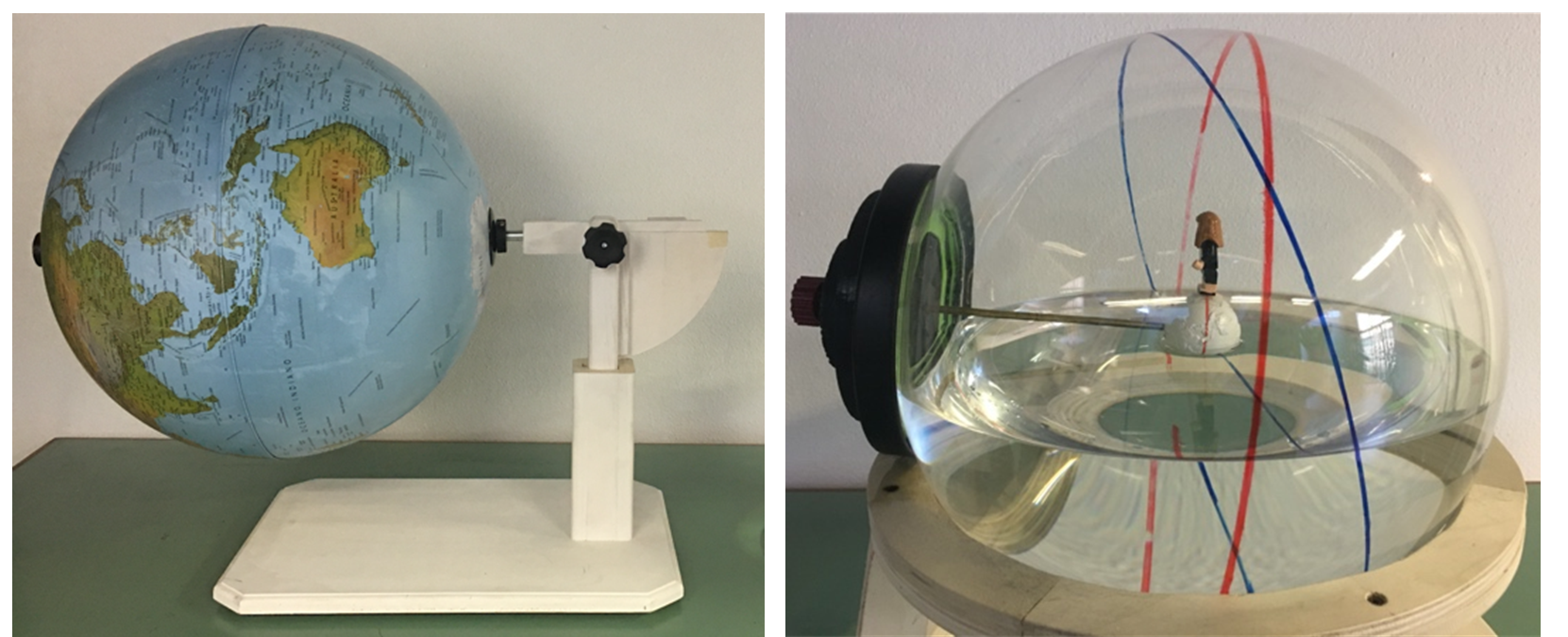}% Here is how to import EPS art
\caption{\label{fig:epsart} The 3D model of Earth and the "bottle globe" model of the celestial sphere with the celestial equator and the ecliptic indicated (\cite{2016_Fisher}), created by the Planetario di Modena.}
\end{figure}

\subsubsection{Planetarium presentation}
The planetarium presentation is about the apparent motion of the Sun and stars. It consists of three major parts: the apparent daily motion of the Sun and stars, the apparent annual motion of the Sun and the observer's location on the spherical Earth. Since not all planetariums have a digital projection system, the presentation only focuses on the geocentric point of view. Each time an element of the Sun's apparent motion is explained and shown, immediately afterwards it is explained and shown for the stars. In this way, we hope to highlight the similarities and differences between the Sun and the stars. 
To keep the students' attention, the presentation lasts maximum 1 hour and interactivity is stimulated by regularly asking questions to the audience.

\begin{figure}[h] % de h staat voor "hier"
\centering
\includegraphics[width=8.5cm]{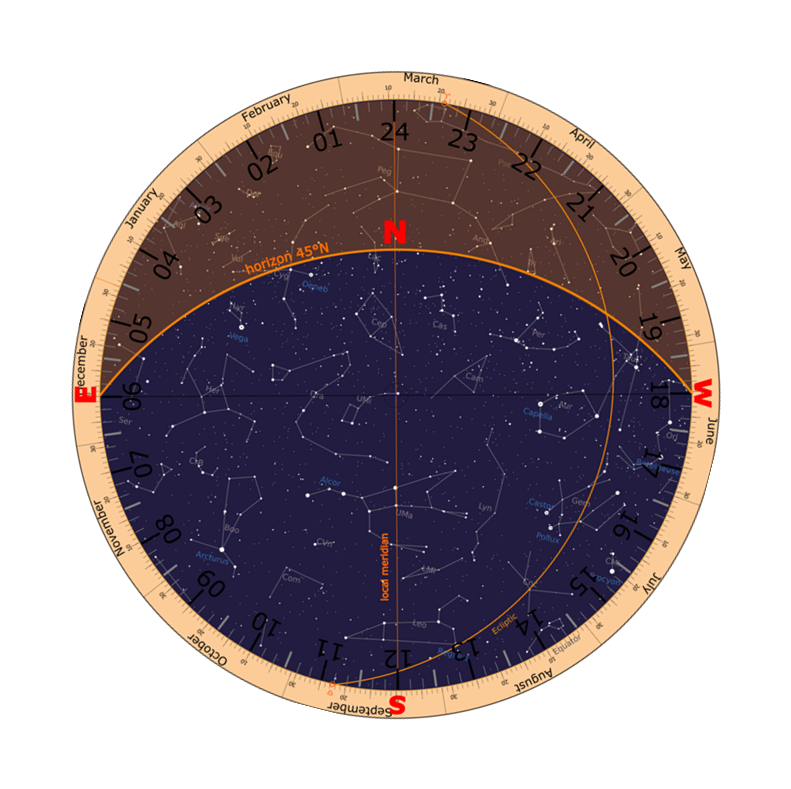}% Here is how to import EPS art
\caption{\label{fig:epsart} The planisphere as used during the planetarium workshop, created by the Planetario di Modena.}
\end{figure}

\subsubsection{Planetarium workshop 2: The planisphere}
The second planetarium workshop is about the use of the planisphere, which is a 2D representation of the celestial sphere. Special about the TASTE planisphere, developed by the Planetaria di Modena (See Figure 3), is that it has two sides: one for the northern and one for the southern hemisphere. On each side it consists of two rotating discs: a disc with the star map, and a disc with the observer's horizon, depending on the observer's latitude. By rotating the horizon disc, students visualize the daily motion of the stars from east to west. They observe that the times of star rise and star set change throughout the year. Through exercises they realize that for a specific location on Earth some stars never set, and other stars never rise. During the workshop students are stimulated to use their smartphone with the "Heavens above" app (developed by Chris Peat) installed and compare the visualisations of the planisphere with the images of the smartphone app.

\begin{figure}[h] % de h staat voor "hier"
\centering
\includegraphics[width=3cm]{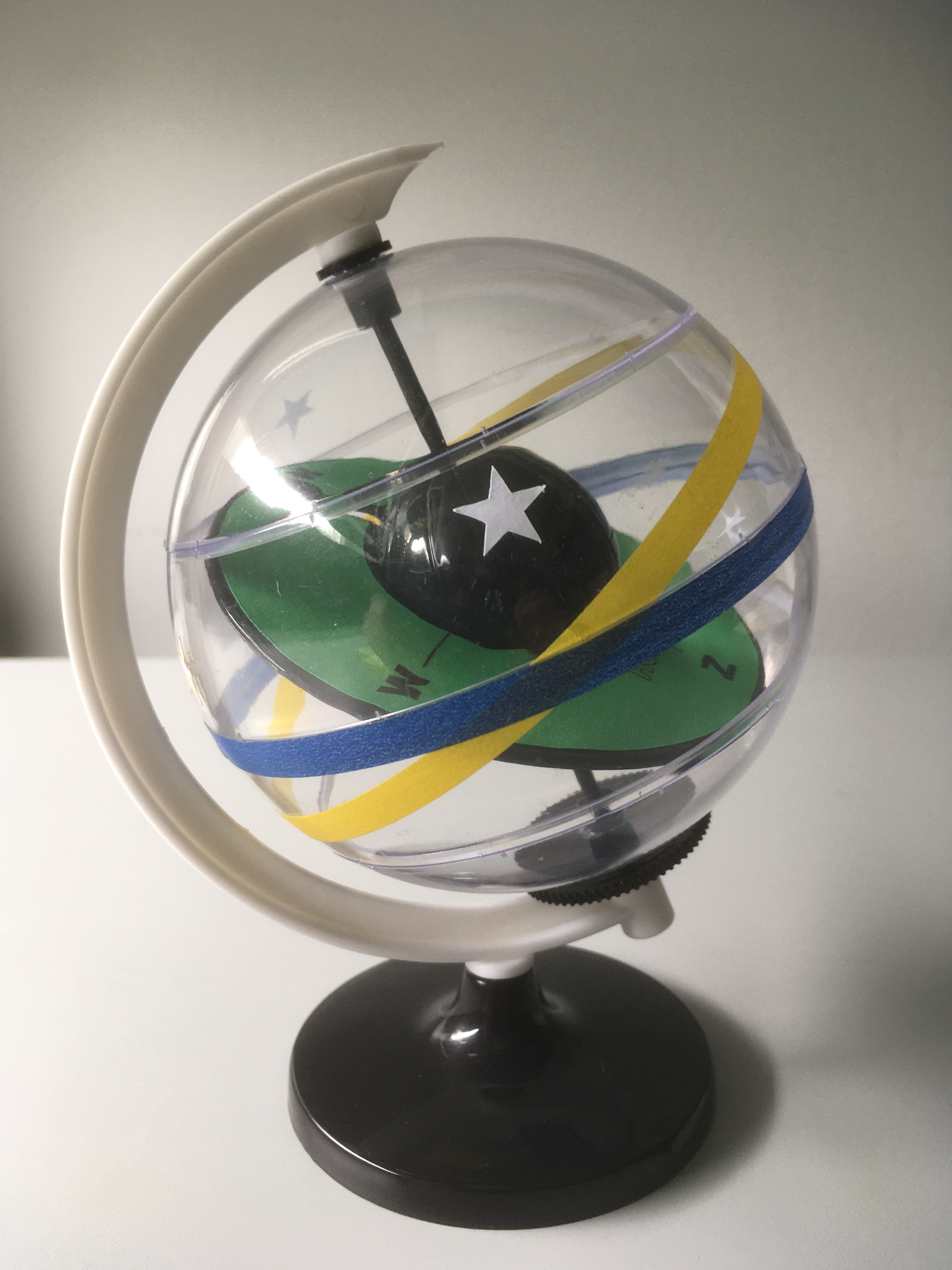}% Here is how to import EPS art
\caption{\label{fig:epsart} The 3D model of the celestial sphere, as used by the Belgian students.}
\end{figure}

\subsection{\label{sec:level2}Two-treatment study}
Through the sequence of the different activities described above, students learn to look at the different elements of the apparent motion of the Sun and stars (rise, culmination point, set, etc.) from different frames of reference and with different representations of the Earth and the celestial sphere. Through this repetition, we encourage students to increase their understanding of these apparent motions.
To test the extent to which the planetarium activities lead to learning gains compared to students who do not visit a planetarium, we have set up a two-treatment study. The two groups are:
\begin{enumerate}
		\item Group 1: The Belgian students followed only the learning module at school (see Table 2).
              \item Group 2: The German, Greek and Italian students followed the full sequence of activities (see Table 3) in collaboration with a planetarium in their school's neighbourhood.
	\end{enumerate}

To allow the students in group 1 to also physically engage with the 3D model of the celestial sphere, some exercises were added to the Belgian version of the learning module in which students used the 3D model at school, while the others used it at the planetarium. A model was chosen that could be used quickly and easily in an ordinary classroom. It was a restyled version of a commercially available tool (see Figure 4). 

\begin{table*}
\caption{\label{tab:table1}Learning sequence for treatment group 1}
\centering
    \begin{tabular}{p{5cm} p{3cm} c p{3cm}}\hline
         & Location & Duration (minutes)& Leader\\ \hline
        Pretest &  School& 20& Teacher \\
        Learning module with two units & School & 80 & Teacher\\
        Posttest & School & 20& Teacher\\  \hline
    \end{tabular}

\end{table*}

\begin{table*}
\caption{\label{tab:table2}Learning sequence for treatment group 2}
\centering
    \begin{tabular}{p{5cm} p{3cm} c p{3cm}}\hline
         & Location & Duration (minutes)& Leader\\ \hline
        Pretest &  School& 20& Teacher \\
        Learning module with two units & School & 80 & Teacher\\
        Workshop 1 & Planetarium site & 80 & Planetarium staff \\
        Planetarium presentation & Dome & 60& Planetarium staff\\
        Workshop 2 & Planetarium site & 80& Planetarium staff\\
        Posttest & Planetarium site & 20& Planetarium staff\\  \hline
    \end{tabular}

\end{table*}

\subsection{\label{sec:level2}Measurement instrument}
The AMoSS test with 12 multiple choice questions focuses on distinctions between different aspects of the apparent motion of the Sun and stars (See Table 4). For each question about the Sun, there is a parallel question about the stars. For an example, we refer to Figure 5. For more information on the test, we refer to a previous manuscript (\cite{2022_Bekaert2}). We used the translated versions of the test in the different languages of the participating students (Dutch, German, Greek and Italian) and adapted the questions to the local situation so that the mentioned cities were familiar for the students (e.g. Brussels, Heidelberg, Thessaloniki and Modena). We also adapted the numerical values for culmination height to correspond correctly to each participating location's situation (see Appendix A).

\begin{table*}
\caption{\label{tab:table3}Framework of the AMoSS test: similarities and differences between the apparent motion of the Sun and stars}
%\begin{ruledtabular}
\renewcommand{\arraystretch}{1.3}
\begin{tabular}{p{0.3cm}p{0.3cm}p{6.5cm}p{0.3cm}p{0.3cm}p{6.5cm}}\hline
(I)&\multicolumn{2}{l}{Apparent motion of the Sun}&(II)&\multicolumn{2}{l}{Apparent motion of a star}
\\ \hline
 &(A)&Daily Sun position changes: Sun's path. (Question I.A)&&(A)&Nightly star position changes: star trail. (Question II.A)\\
 &(B)&Sun culmination changes during a year. (Question I.B)&&(B)&Star culmination does not change during a year. (Question II.B)  \\
 &(C)&Sunrise and sunset position change during a year. (Question I.C)&&(C)&Star-rise and star-set position do not change during a year. (Question II.C)  \\
 &(D)&Sun culmination depends on observer position. (Question I.D)&&(D)&Star culmination depends on observer position. (Question II.D) \\
 &(E)&Sunrise and sunset position depend on observer position. (Question I.E)&&(E)&Star-rise and star-set position depend on observer position. (Question II.E) \\
 \hline
 (III)&\multicolumn{2}{p{7cm}}{Seasons: colder and warmer periods on a specific location during a year, due to Earth's revolution. (Question III)}&(IV)&\multicolumn{2}{p{7cm}}{Sky map changes on a specific location during a year, due to Earth's revolution. (Question IV)}
 \\ \hline
\end{tabular}
%\end{ruledtabular}
\end{table*}

\begin{figure}[h] % de h staat voor "hier"
\includegraphics[width=8.5cm]{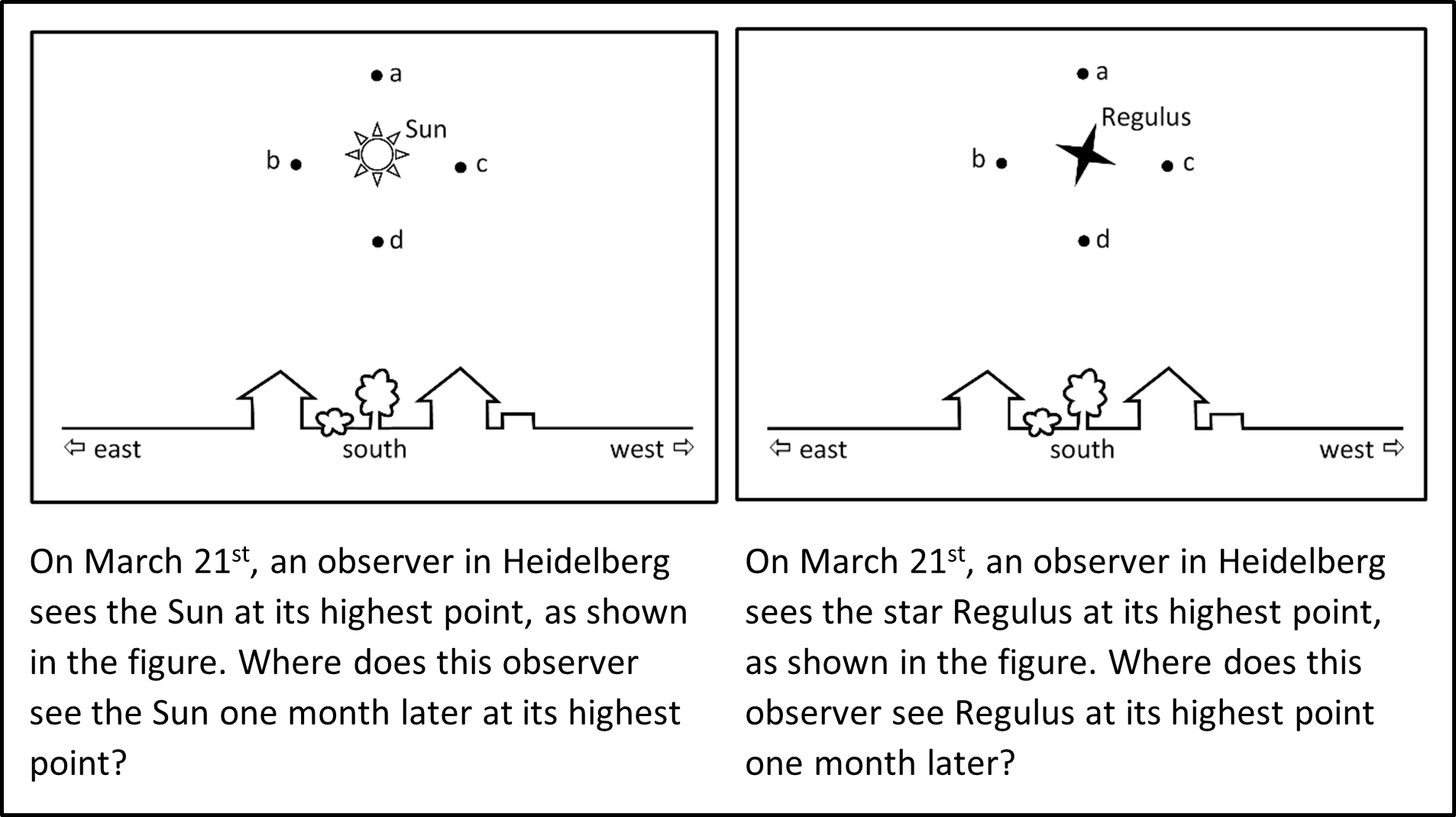}% Here is how to import EPS art
\caption{\label{fig:epsart} Question I.B and II.B of the AMoSS Test (\cite{2020_Bekaert}) for German students.}
\end{figure}

 We administered the AMoSS test as a pretest to secondary school students (13-17 years old) of the four participating countries during a science lesson at school, before the learning activity at school. The students were not specially prepared to the test. 

The group 1 students took the AMoSS test again as a posttest, immediately after the lesson activity at school, since they were not able to follow planetarium workshops, nor a planetarium presentation in the dome. The group 2 students went to a planetarium in their schools' neighbourhood where they participated in a first workshop about the 3D-model of the Earth and the celestial sphere, then followed a planetarium presentation in the dome, and finally joined a second workshop about the use of a planisphere. After the second workshop they took the AMoSS test as their posttest, at the planetarium site.

We asked the twelve questions in a random order. To exclude a bias in the results due to this order, we created two different series, each with an identical set of questions, but in a different order.

A protocol was written and explained to all teachers and planetarium staff, so that the test administration and intervention were implemented in the same way in all schools and planetariums. The test lasted 20 minutes. Unlike in the previous study, we did not ask for explanations, due to time constraints.

For the analysis of the multiple choice answers a score of 1 was given if the correct alternative was chosen and 0 if an incorrect alternative was chosen or if no answer was given.

Using an independent samples t-test, we examined whether the two groups were significantly different in the pretest. With a single factor ANCOVA, in which we take results from the pretest as a covariate in order to control for existing differences at the pretest, we checked if the two groups differ from each other in the posttest. To look for possible differences between the means of the pretest and posttest, we performed a paired samples t-test.

Other than only looking at the simple statistics of the students' answers, we also applied a Latent Class Analysis (LCA) on the multiple choice answers to look for possible patterns in the students' answers. LCA is an appropriate statistical tool to uncover patterns in the checked alternatives \cmmnt{(REF. artikel 359 p. 288 en 289)}(\cite{2020_Weller}). We used the LCA functionality of software Mplus and entered a data file with the answers of the multiple choice questions (a, b, c, …) of the students on 10 out of 12 questions of pretest and posttest. We excluded questions III and IV about seasons and sky map changes throughout the year, to be in line with our prior research (\cite{2020_Bekaert, 2022_Bekaert2}). We made one data file with all student responses, to be able to map any possible evolution between pretest and posttest.

\section{Findings}

\subsection{\label{sec:level2}Descriptive results of the multiple choice answers}

We present the mean test score of the pretest and posttest, the median and the standard deviation for the two treatment groups in Table 5. We distinguish between the Sun and the star questions. We present the results per country in Appendix C.

\begin{table*}
\caption{\label{tab:table3}Student scores on the AMoSS test for group 1 (\textit{N} = 90) and group 2 (\textit{N} = 172). If the difference between the pretest and posttest is significant (p $<$ 0.05), this is marked by an asterisk\textsuperscript{*}.}
%\begin{ruledtabular} 
\renewcommand{\arraystretch}{1.3}
\begin{tabular}{p{2.8cm} p{0.9cm} p{0.9cm} p{0.9cm} p{0.4cm} p{0.9cm} p{0.9cm} p{0.9cm} p{0.4cm} p{0.9cm} p{0.9cm} p{0.9cm}} \hline
        & \multicolumn{3}{c}{Sun questions}    &    & \multicolumn{3}{c}{Star questions} & &\multicolumn{3}{c}{All questions }\\
&PRE &POST &GAIN & &PRE &POST &GAIN &   &PRE &POST &GAIN\textbf{}
\\ \hline
\multicolumn{4}{l}{GROUP 1 (\textit{N} = 90)}
\\
Mean &54 \%\textsuperscript{*}  &66 \%\textsuperscript{*}  &+12 \% & &30 \%\textsuperscript{*} &47 \%\textsuperscript{*} &+17 \% & &42\%\textsuperscript{*} &56\%\textsuperscript{*} &+14 \%\\
Median  &67 \%    &67 \% & & &33 \%            &50 \%  & &   &42 \% &58 \% \\
Standard deviation   &25 \%    &22 \% & & &18 \%    &23 \%  & &   &17 \% &18 \%
\\ \hline
\multicolumn{4}{l}{GROUP 2 (\textit{N} = 172)}
\\
Mean &33 \%\textsuperscript{*}  &49 \%\textsuperscript{*} &+16 \% & &22 \%\textsuperscript{*} &33 \%\textsuperscript{*} &+11 \% & &28\%\textsuperscript{*} &41\%\textsuperscript{*} &+13 \% \\
Median  &33 \%    &50 \% & & &17 \%            &33 \%   & &  &25 \% &42 \% \\
Standard deviation   &22 \%    &23 \% & & &17 \%    &21 \%   & &  &16 \% &18 \%  \\ \hline
\end{tabular}
%\end{ruledtabular}
\end{table*}

Table 5 makes clear that both pretest and posttest scores are low and the Sun-related questions are answered somewhat better than the star-related questions. A paired-samples t-test was conducted to compare the mean scores of the pretest and posttest. An asterisk\textsuperscript{*} in the table indicates a significant difference ($p < 0.05$) in a paired samples t-test of the pre- and posttest scores. An independent samples t-test also shows that the overall pretest scores in the two groups are significantly different ($p < 0.001$) for both the Sun ($p < 0.001$) and star questions ($p < 0.05$). A single factor ANCOVA indicates that all posttest results were also significantly different for the two groups ($p < 0.001$), taking the pretest score as a covariate.  

The overall learning gain in the two treatment groups is very similar, 14\% and 13\% respectively. An independent samples t-test reveals that learning gains in the two groups do not differ significantly ($p > 0.05$). In the second group, the learning gain for the Sun questions is larger than the star questions, while in the first group the strongest learning gain is achieved for the star questions.

To examine how this average learning gain is reflected in the different questions of the pretest and posttest, Figure 6 shows details at the question level. Again, we present the results per treatment group. These graphs confirm that in both groups in the posttest students answer most Sun questions better than in the pretest, but we see important differences between questions and between groups. For example, among group 2 students, we see the strongest improvement on the question on the daily motion of stars (Question II.A), while the group 1 students barely perform better on the posttest, compared with the pretest. On the other hand, only in group 1 the students improve their knowledge about the star trails throughout the year (Question II.B and II.C). It is striking that group 2, that went to the planetarium, answered these questions about the stars worse in the posttest than group 1, that only followed the activity at school.
Concerning the questions about the observers' position (Questions D and E), in both groups following the activities have little impact on students' scores on these questions. 
Finally, the progress on the questions about the seasons (Question III) and the change of the sky map (Question IV) throughout the year is remarkable in both groups. 

\begin{figure*}%[b!]  %% Add a [b!] if you prefer the wide image to be at the bottm of the page
\centering
\includegraphics[width=1.0\textwidth]{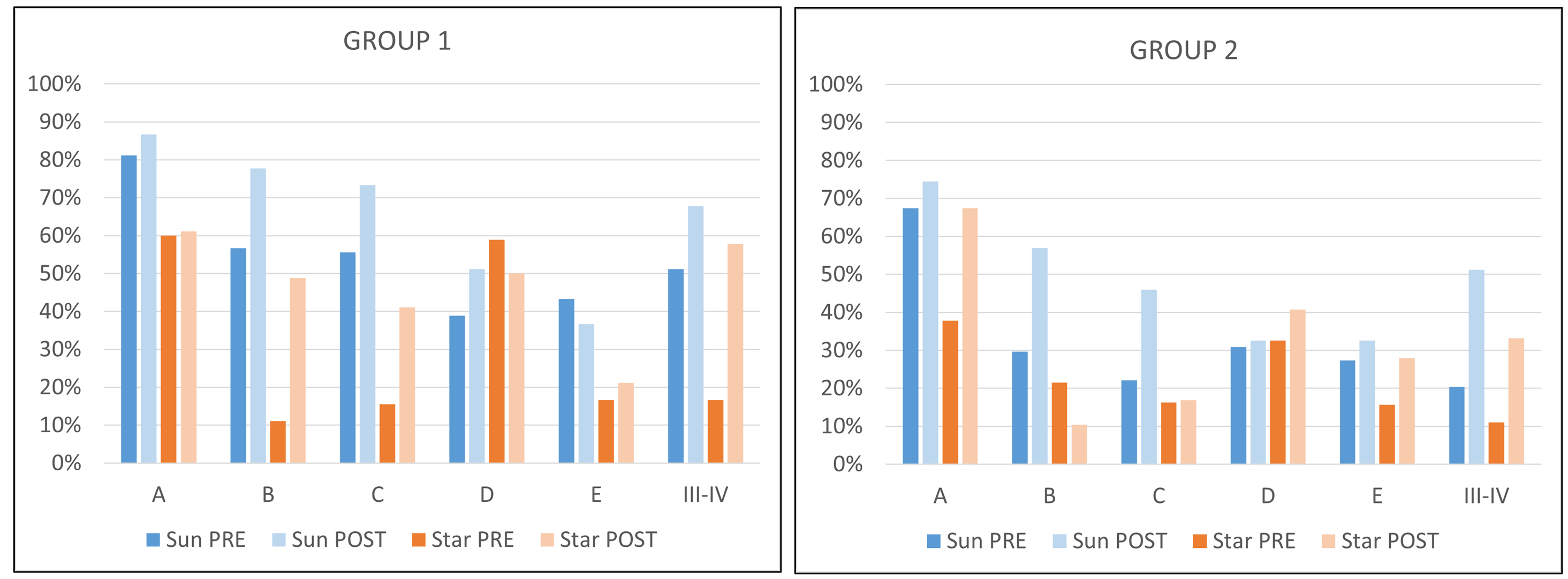}
\caption{Percentage of students with a correct answer on pretest and posttest, ordered per question
}\label{fig:example:wide}
\end{figure*}

To check for the eventuality that the learning module and planetarium activities might have had a negative impact on some student's scores, we examined for each student and for each question whether the student's original correct answer in the pretest changes to a wrong answer in the posttest. Figure 7 gives an overview of this evolution. We see that in both groups for each question, a limited group of students transform their correct answer into a wrong one (indicated by pink colour in Figure 7). It is striking how most group 2 students stick to their wrong answer on questions II.B and II.C or shift to a wrong one, while the group 1 students tend to shift towards a correct answer.

\begin{figure*}%[b!]  %% Add a [b!] if you prefer the wide image to be at the bottm of the page
\centering
\includegraphics[width=1.0\textwidth]{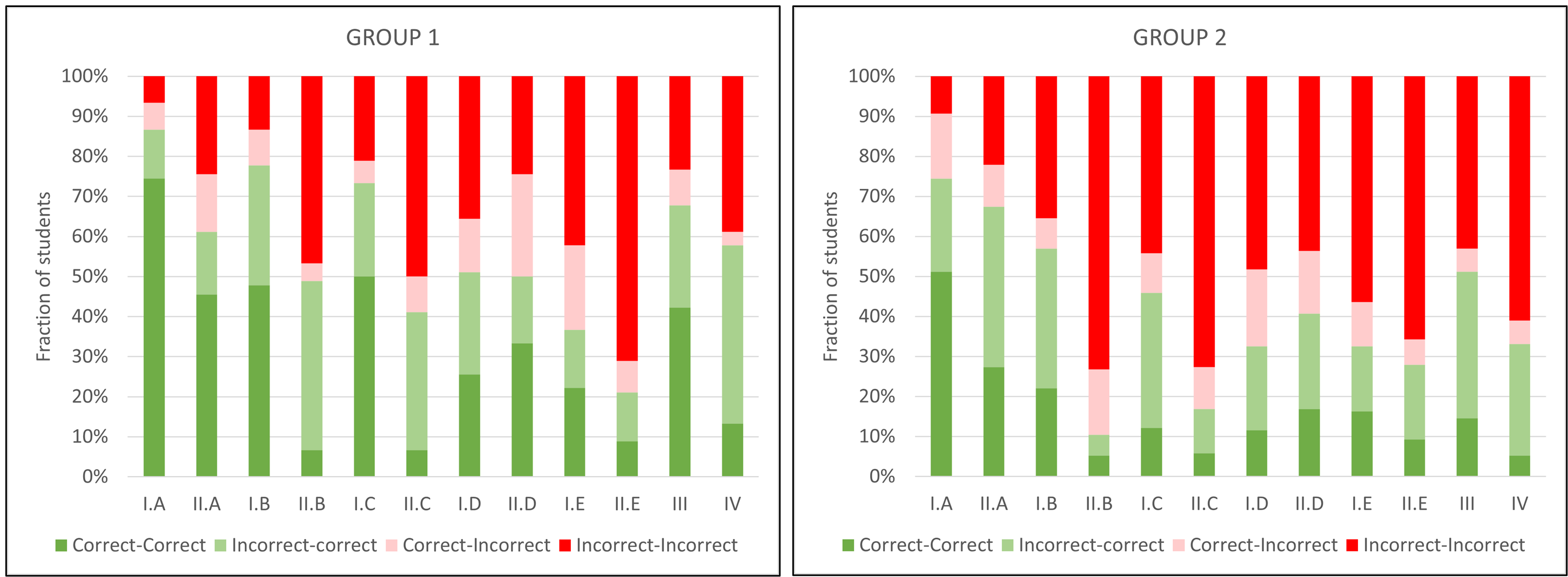}
\caption{Change of correctness between pretest and posttest, ordered per question}\label{fig:example:wide}
\end{figure*}

\subsection{\label{sec:level2}Latent class analysis}
\subsubsection{Choosing the number of classes}
In the first step of a latent class analysis, we determined the number of classes.  We calculated two fit indices to decide which model suits our data best: the Akaike Information Criterion (AIC) \cmmnt{(AIC; Akaike, 1974 thesis Elien Van Luydt p. 51 – artikel 361)}(\cite{1974_Akaike}) and the Bayesian Information Criterion (BIC) \cmmnt{; Schwarz, 1978 – artikel 362)}(\cite{1978_Schwarz}). As a rule, the model with the lowest AIC and BIC values corresponds to the one with the best model fit. \cmmnt{(Bron: google books Environmental Policy Change in Emerging Market Democracies of Google book Confirmatory Factor Analysis for Applied Research, Second Edition, p. 156)}Here the model with two and seven classes (see Table 6), have respectively the lowest BIC and AIC. From a theoretical interpretation point of view, the seven class solution is most preferable. \cmmnt{(Norusis, 2011, Norusis, M.J. (2011). IBM SPSS Statistics 19 Procedures Companion. Addison Wesley, TX.)}A description with two classes is rather uninformative because it basically provides a class with students who score high and a class with students who score low on the test. A description with seven classes is substantially more informative as it provides groups that show qualitatively different answering patterns which indicate different mental models, and the seven classes that were obtained moreover are consistent with what was found in previous studies. Therefore, we choose the seven-class solution.

\begin{table}[h]
\caption{\label{tab:table3}AIC and BIC values for different number of classes}
\centering
%\begin{ruledtabular}
\renewcommand{\arraystretch}{1.3}
\begin{tabular}{c c c} \hline
Number of classes       & AIC        & BIC   \\\hline
1       &15178.278        &15399.876\\
2       &14783.618        &\underline{15231.074}\\
3       &14688.583        &15361.899\\
4       &14609.725       &15508.900\\
5       &14562.565       &15687.598\\
6       &14534.299         &15885.192\\
7       &\underline{14526.711}        &16103.463\\
8       &14539.039         &16341.65\\\hline
\end{tabular}
%\end{ruledtabular}
\end{table}

\subsubsection{Description of the classes}
In the output of the latent class analysis, conducted by the software Mplus, the size of the seven classes is indicated by the probability that a respondent belongs to a certain class. These posterior probabilities are calculated based on the respondent answers on the multiple choice questions. Table 7 shows these probabilities for the different classes, distinguishing between pretest and posttest. 

\begin{figure*}%[b!]  %% Add a [b!] if you prefer the wide image to be at the bottm of the page
\centering
\includegraphics[width=1.0\textwidth]{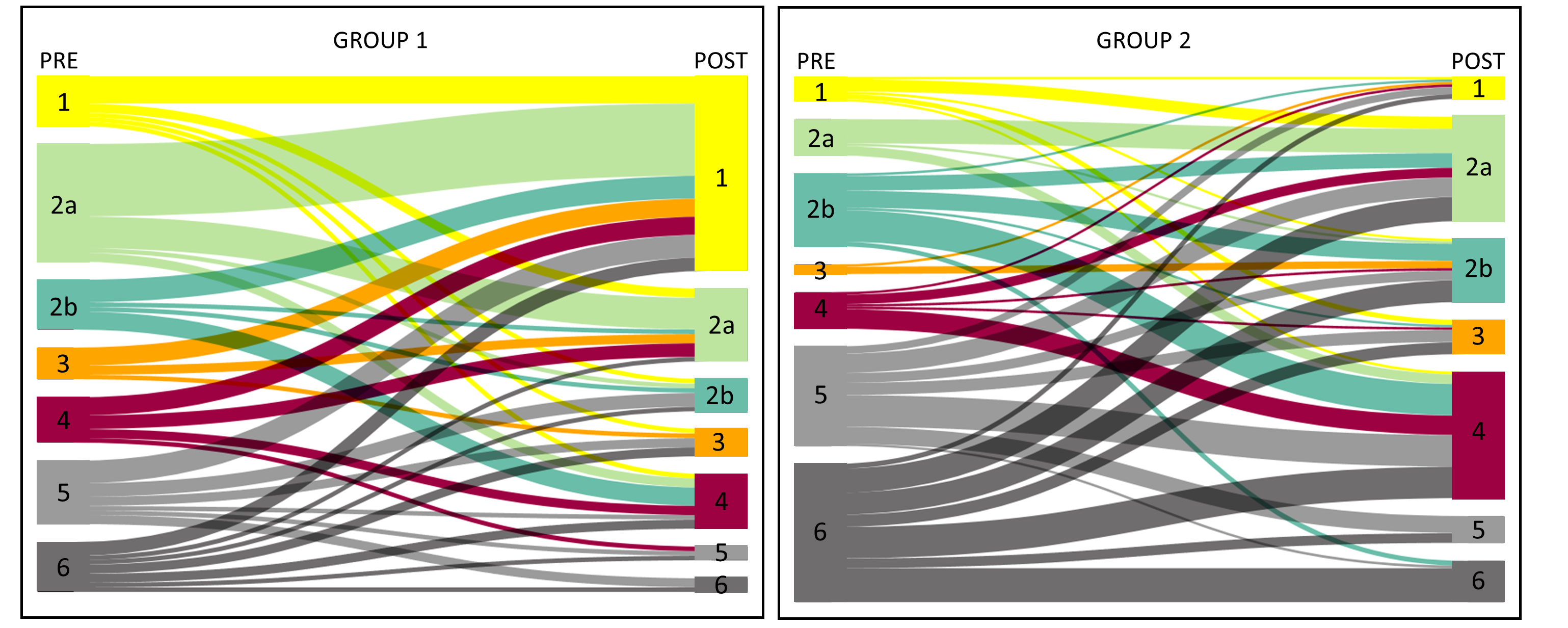}
\caption{Evolution of class membership between pretest and posttest, presented by country. The higher the class is positioned in the figure, the better the student scores on the AMoSS test.
}\label{fig:example:wide}
\end{figure*}

\begin{table}[]
\caption{\label{tab:table3}Predicted memberships of the latent classes in pretest and posttest for all students (\textit{N} = 262)}
\centering
%\begin{ruledtabular}
\renewcommand{\arraystretch}{1.1}
\begin{tabular}{p{1.1cm} c c} \hline
    & PRE &POST  \\\hline
Class 1 &8\% &20 \%\\
Class 2a &16\% &22 \%\\
Class 2b &16\% &13 \%\\
Class 3 &4\% &8 \%\\
Class 4 &9\% &24 \%\\
Class 5 &21\% &5 \%\\
Class 6 &26\% &8 \%\\
\hline
\end{tabular}
%\end{ruledtabular}
\end{table}

To reveal the differences between classes, for each question the software Mplus generates a table with the probabilities that a certain answer is given by a member of a certain class. Table 8 presents an example for question II.B and can be interpreted as follows: there is a 3 \% chance of a respondent in latent class 1 answering alternative a to the multiple choice question, a 0 \% chance of answering alternative b, a 8 \% chance of answering alternative c, … The "x" in the table represents a blank answer. In Appendix B this table is presented for the other questions.

\begin{table}[]
\caption{\label{tab:table3}Example of the output of software Mplus: Probability per class for each alternative of question II.B (in \%).}
\centering
%\begin{ruledtabular}
\renewcommand{\arraystretch}{1.1}
\begin{tabular}{p{1.1cm} c c c c c c c} \hline
Class       & a     &b      &c      &d      &e      &f      &x     \\\hline
Class 1    &3  &0   &8  &6  &62  &22     &0 \\
Class 2a    &60  &1  &18  &19 &0  &4    &0 \\
Class 2b    &27 &40  &0  &6 &20 &7   &0 \\
Class 3     &33  &11  &43 &0  &8   &5    &0  \\
Class 4     &7   &8   &45   &26   &13   &0    &1  \\ 
Class 5     &10   &5   &14   &1  &46  &24   &0 \\  
Class 6     &4  &0   &19   &6   &6   &65   &0  \\ \hline
\end{tabular}
%\end{ruledtabular}
\end{table}

Based on these tables we describe the profiles of the students in the seven classes in Table 9. In this table we also make explicit which progress the students can make to improve their understanding of the apparent motion of the Sun and stars.

\begin{table*}
\caption{\label{tab:table3}Description of the classes and possible learning opportunities for the students in each class. The higher the class number, the more learning opportunities the students have.}
\centering
%\begin{ruledtabular}
\renewcommand{\arraystretch}{1.1}
\begin{tabular}{p{1.1cm} p{7cm}  p{7cm} } \hline
Class       & Description     &Learning opportunities  \\\hline
Class 1    &The students in class 1 answer all questions correctly.  & \\ \hline
Class 2    &The students in class 2 answer all questions about the Sun correctly. Also the star questions are answered mostly correctly, except the questions about the change of the star trails throughout the year (question II.B and question II.C). Concerning these two questions this class is subdivided in two subclasses.  &Students need to understand that
(1) the Sun is a specific star for us because Earth orbits the Sun; (2) because of the tilt of the Earth's axis relative to the ecliptic plane and because of the Earth's revolution around the Sun, the Sun's path changes from day to day; (3) the star trail of a specific star (other than the Sun) stays fixed throughout the year.
\\ 
    &\textbf{Class 2a:} The students think that the stars seem to behave exactly the same as the Sun. For both the Sun and the stars, in summer the path of the apparent motion is higher and wider than in winter.    & \\
    &\textbf{Class 2b:} The students are confused about how the star trails depend on the time of the year. They don't know that the stars' culmination height and positions of rise and set stay fixed throughout the year.  & \\ \hline
Class 3    &Only the questions about the apparent daily motion of the Sun and stars and about the apparent annual motion of the Sun, are answered correctly. This means most star question are answered incorrectly. Students do not know how the culmination height depends on the latitude of the observer.  &In addition to the above learning opportunities students need to understand how
(4) the observer's position affects the way the Sun's path and the star trails are observed. Both the culmination height and the position of rise and set of the Sun and stars change as the observer's latitude changes.   \\ \hline
Class 4    &These students know that the Sun and stars appear to move from east to west during the day and night and that the Sun's culmination height increases during spring. All other questions are answered incorrectly. They think that the culmination height is proportional to the latitude of the observer.  &In addition to the above learning opportunities students need to understand that 
(5) the position of rise and set of the Sun and stars is linked to the culmination height of the Sun and stars. A different culmination height, results in different positions or rise and set.\\ \hline
Class 5    &Concerning the daily apparent motion, the students in class 5 know that the Sun and the stars appear to move from east to west. Concerning the annual motion these students think that the Sun's path and the star trails do not change throughout the year. When the observers' latitude changes, they think that the positions of Sunrise and Sunset do not change and that the star trails do not change at all. On the other hand, as the students in class 4, they think that the culmination height is proportional to the latitude of the observer.  &In addition to the above learning opportunities students need to understand that 
(6) the Sun's culmination height changes from day to day. \\ \hline
Class 6    &This group is characterized by the fact that students answer almost all questions with "I don't know", except the first question about the Sun's daily motion.  &In addition to the above learning opportunities students need to understand that 
(7) facing the south the stars move in the sky from east to west, due to Earth's rotation around its axis. \\ \hline
\end{tabular}
%\end{ruledtabular}
\end{table*}

%\vspace{5mm} %5mm vertical space

Table 7 shows that students switch classes by following the learning module and planetarium activities. In Figure 8, we disentangle this evolution in detail per group. We see that most students change classes between pretest and posttest. In the first treatment group that did not attend a planetarium presentation, there is a notable shift to class 1, while in the other group that followed two planetarium workshops and a planetarium presentation, there is a major shift to class 2 and 4.

\section{Discussion}

In the context of the European Erasmus+ project TASTE, we investigated the extent to which a sequence of activities at school and the planetarium support secondary school students to improve their insight in the apparent motion of the Sun and stars. These activities were developed by teachers in collaboration with planetarium staff. The focus is on a step-by-step application of the celestial sphere model and practising its use through hands-on activities.

The study is a two-treatment study with a pretest posttest design. As a pretest, we have administered the AMoSS test (\cite{2020_Bekaert, 2022_Bekaert2}) with 12 multiple choice questions to 262 students (13-17 years old), coming from four different European countries (Belgium, Germany, Greece and Italy). All students started the sequence of activities with a 80-minutes learning module at school. The first treatment group, the Belgian students, only took this learning module and then completed the AMoSS test again as a posttest. The second treatment group, the German, Greek and Italian students, went on a field trip to the planetarium and attended two workshops and a planetarium presentation in the dome. After these planetarium activities they also took the AMoSS test as a posttest.
The systematic design of the AMoSS test, with six Sun-related questions and six parallel star-related questions,  allows us to compare students’ learning gains of the Sun’s apparent motion with the stars' apparent motion. 
With the Latent Class Analysis technique, we examined the extent to which the sequence of activities affect the mental model that students use when answering questions on the Sun and stars.

% \subsection{\label{sec:level2}Students' results on the AMoSS test}
The pretest results of the second treatment group give a mean score of 28\% (see Table 5). The students of the first treatment group perform better with an average of 42\%. The difference between the two groups is significant ($p < 0.001$). This may be due to the fact that the average age of the first group students is a bit higher, which means that these students have already taken some more astronomy classes at school. For both groups, the mean score of the Sun-related questions (33\% and 54\%) is higher than for the star-related questions (22\% and 30\%). This confirms previous studies that also report a better student view on the Sun's apparent motion than the stars' apparent motion \cmmnt{(ART 1) REF [14,15,17,18])} (\cite{1993_Mant,2009_Plummer,1994_Vosniadou,2022_Bekaert,2022_Bekaert2}). This should not surprise us as we witness the Sun's apparent motion during the day more often and more easily than we observe the night sky. Moreover, in school curricula, the study of the Sun is covered more prominently than the study of the stars. 

Comparing the mean scores of the posttest for the two groups (56\% and 41\%), the first treatment group performs also a bit better than the second group. Taking the pretest score as a covariate, a single factor ANCOVA indicates that this difference is significant ($p < 0.001$). The learning gain from pretest to posttest is significant in both groups and is similar (14\% and 13\%). Although we see a significant improvement in scores, we consider the learning effect as rather limited. In addition, students who only did the learning module at school made the same progress as students who took both the learning module at school and a trip to the planetarium. Like \cite{1970_Reed}, we can conclude that a teaching activity at school where students used a 3D model of the celestial sphere provides as much of a learning benefit as a planetarium activity, wherein this model is used. Moreover, we think that the learning gains of students who attended the planetarium activities can mainly be attributed to the hands-on character of the workshops, where students learned and practiced to use 3D models. Indeed, the planetarium presentation in the dome was fairly classically structured because it had to be able to be presented in both an analogue and a digital planetarium and so did not focus on perspective switching. More research is needed to clarify this.

While we see few differences in learning gains between the different treatment groups, there are clear differences when we look at learning gains per question, in particular at star-questions (see Figure 6). In relation to the daily motion of stars (question II.A), there is hardly any learning gain for the first group, while the results of the second group improve significantly. We think this can be attributed to the fact that the first group did not attend a planetarium presentation. They were not immersed in the apparent rotation of the night sky, which can be beautifully simulated in a planetarium dome, regardless of whether this is in an analogue or digital setting. We agree with the literature \cmmnt{(REF. artikel (420) – artikel (421))}(\cite{2016_Yu}) that being immersed stimulates the students to reason about the concepts taught. 

While this worked well for the daily motion of stars, we see the opposite effect for the star trails throughout the year (questions II.B and II.C). The group 1 students who only followed the learning module at school show a clear learning gain on these questions, while in the other group it is a rather small gain and even loss. For these questions, there was a large number of group 1 students who converted a wrong answer in the pretest into a correct answer in the posttest, while this was not the case for the other students (see Figure 7). In accordance with constructivism, each student attempts to develop his/her understanding based on the learning material provided in interaction with his/her prior knowledge. This may result in students developing alternative ways of thinking as a result of instruction. Because both treatment groups followed the same learning module at school, the difference between the two groups cannot be immediately explained, unless it is due to the fact that we added some exercises to the learning module of the first group. Since this group did not attend the workshop on 3D models at the planetarium, we supplemented the learning module at school with exercises on how to use the 3D model of the celestial sphere. They could mark a Sun or star on the plastic sphere and rotate the sphere to simulate the apparent motion of Sun and stars. The second group also made similar exercises, but in a separate workshop at the planetarium and with a different 3D model where the rotational motion was more difficult to simulate.  Concerning star trails throughout the year, at first sight, an integrated approach, where 3D models are also used in the classroom, gives better results then combining a planetarium presentation in the dome with workshops. A second possible explanation could be that the students in the first group are on average a bit older and consequently have already been taught more about the motion of celestial bodies. This may have helped them to make progress. More research is needed to explain the difference in results between the two treatment groups. 

With regard to the spherical shape of Earth and how this affects the way an observer at different latitudes perceives the apparent motion, neither taking the class at school nor following a planetarium presentation provides a clear learning gain for the Sun and star questions about this topic (Question D and E). Although the concept of latitude is explained and practiced in the learning module based on the position of Polaris in the sky for observers at different places on Earth (North Pole, equator, ..), there was no separate part specifically about this. We have deliberately chosen not to treat this as a separate chapter due to time constraints. Because science teachers are under time pressure to complete the regular curriculum within the allocated lesson time, it was stipulated that the learning module had to be completed in 80 minutes. As we found in our previous research (\cite{2022_Bekaert}) that students do not automatically transfer knowledge about Earth's rotation and revolution to the apparent motion of the celestial bodies, neither do they for the spherical shape of Earth. If we want students to improve their understanding, we will have to teach specifically about this topic and will have to provide specific exercises. 
The fact that the students who followed the planetarium presentation also did not perform better on these questions, shows that also in the planetarium presentation extra attention must be paid to this. Because the presentation had to be able to be presented in both an analog and a digital system, only a geocentric point of view could be demonstrated. As discussed in the literature \cmmnt{(REF. ART 2 – zie background)}(\cite{2010_Plummer,2011_Lopresto,2015_Testa,2015_Yu}), it seems to be essential that an allocentric perspective can also be shown during the presentation in the dome.

With the last two questions about the seasons and the change of the star sky throughout the year (Question III and IV), both groups improve in the posttest. With both questions, we see that in both groups many students convert a wrong answer in the pretest into a correct answer in the posttest. We attribute this to the fact that these questions are about items literally covered during the learning module and in the planetarium activities. Students who have picked up and remembered this have been able to score well on these questions.

\subsection{\label{sec:level2}Students' mental models}
As in an earlier study (\cite{2022_Bekaert2}), we performed a latent class analysis (LCA) on the answers to the multiple choice questions to see which patterns could be detected in students' answers. The number of students who answered all questions correct grew from 8 \% in the pretest to 20 \% in the posttest (see Table 7). This increase indicates that for a certain group of students, the activities have resulted in a better understanding of the apparent motion of the Sun and stars. This number is significantly higher in the first treatment group than in the second treatment group (see Figure 8). It is notable that in this first group from all classes in the pretest, students move up to this class in the posttest. These students, who only followed the learning module at school, in a sense perform better than those who went to the planetarium. 
For the students in the second treatment group, who took a trip to the planetarium, we see three main shifts from pretest to posttest. First, there is a main shift to the second class of students, who answer all Sun questions correctly, but think that the stars behave exactly as the Sun. They think that like the Sun's path, the star trails get higher and wider in summer, and lower and shorter in winter. Also in the first treatment group, many students belong to this class both in the pretest and posttest. These students form their mental model through analogical thinking (\cite{1987_Collins_BOOK}): being less familiar with the apparent motion of the stars, they copy their knowledge about the Sun to the stars without adaptation. 
Second, the largest shift in the second treatment group is seen towards class 4 with students who think that the culmination height is proportional to the observer's latitude. This shows that the learning materials developed did not pay sufficient attention to the observer's position.
Third, class 6 with students who answer mostly "I don't know", is in the second treatment group strongly represented in the pretest, but this class spreads completely over the other classes in the posttest. 
The way students evolve between pretest and posttest shows that the sequence of the learning module at school, followed by the planetarium activities, mainly improves students' understanding of the Sun, but in relation to the stars, much remains unclear. This confirms what is described in the literature (\cite{2019_Vosniadou}) that transforming an alternative mental model into a correct scientific model is a long and gradual process. Science instruction needs to help students to become aware of their alternative mental models and provide information and training to develop scientific reasoning skills.

\section{Limitations}
Since the AMoSS test was taken in six schools and three planetariums in four different European countries, we have to take into account that the test may not be administered in the same conditions everywhere, along with the fact that the students' age and their curriculum differed from country to country. Despite the fact that the teachers and planetarium staff used the same teaching materials and followed a training, the activities will have been somewhat different in different countries.

As in our first study, the fact that we used a convenience sample in this study is an important limitation: the schools are participating in an Erasmus+ project. The students were selected on a voluntary basis. Moreover, the number of participating students was rather small in some countries.

To check whether the learning gains achieved are also sustainable, it would be worthwhile to organise a retention test a few weeks after the posttest. Within the framework of the Erasmus+ project, this was not possible due to time constraints. 

In further research, it would be interesting to compare the results of different interventions for groups that are more comparable and where a retention test is organized several weeks after the intervention.

%	\section{Conclusions}
\section{Conclusions and implications for astronomy education}

We conclude that - despite substantial and specifically targeted efforts to achieve this goal - a deep understanding of the apparent motion of the Sun and stars is difficult to obtain for the different treatment groups. Overall the Sun-related questions are better answered than the questions about the stars.
The fact that students from two treatment groups show different learning gains on different questions indicates that finding the right approach can make a clear difference in the extent to which students increase their understanding of the apparent motion of the Sun and stars. Following a planetarium presentation in the dome helps to get a better view of the diurnal apparent motion of the stars during the night. Working with a 3D model of the celestial sphere during lessons at school, helps to learn how the Sun's path or the star trails may or may not depend on the time of the year. 
We suggest to use a 3D model of the celestial sphere to explain the apparent celestial motions both in the classroom and during the planetarium presentation in the dome. It will help the students to learn and switch between a geocentric and an allocentric point of view.

\section{Acknowledgements}

The authors would like to thank all students, teachers and staff members of the planetariums and science centres who participated in this study, especially Maria Karakolia, George Bokovos, Eleni Kalaitzidou, Dimitrios Memtsas, Tryfon Toganidis, Jolien Roskams, Lieve Rens, Robin Mommen, Koen Vandevenne, Tom Duchamps, Anke Threimer, Frederic Briend, Samantha Brown, Carolin Liefke, Anna Sippel, Andrea Betti, Marco Parmiggiani, Paola Ferrari, Cinzia Gianaroli, Tiziana Grandi, Rita Scaffidi and Francesca Gherpelli.

This research is funded by the Erasmus+ program of the European Union (2020-1-IT02-KA201-079528). Neither the European Commission nor the project's national funding agency INDIRE are responsible for the content or liable for any losses or damage resulting of the use of these resources.

%% Specify your .bib file name here, without the extension
\bibliography{SLOPEBIBTEX11}

\section{Appendix A: learning materials}
All learning materials (learning module, planetarium workshops, planetarium presentation) are compiled in "TASTE student's book." This is available online via the link \url{https://zenodo.org/records/8288617}. 

The scenario of the planetarium presentation and the abstract for teachers is available via the link \url{https://zenodo.org/records/8297640}. 

The AMoSS test instrument is online available in different languages:
\begin{enumerate}
\item Dutch: \url{https://zenodo.org/records/8305505}
\item English: \url{https://zenodo.org/records/8305505}
\item Greek: \url{https://zenodo.org/records/8305505}
\item German: \url{https://zenodo.org/records/8304813}
\item Italian: \url{https://zenodo.org/records/8304474}
\end{enumerate}

\pagebreak

\section{Appendix B: latent class analysis}

\begin{table}[h]
	\caption{\label{tab:table3}Output of software Mplus for question I.A (in \%)}
 \centering
	%\begin{ruledtabular}
		\renewcommand{\arraystretch}{1.1}
		\begin{tabular}{p{1.1cm} c c c c c c c} \hline
			Class       & a     &b      &c      &d      &e      &f      &x     \\\hline
			Class 1 &4 &4 &91 &0 &1 &0 &0 \\
			Class 2a &9 &0 &91 &0 &0 &0 &0  \\
			Class 2b &0 &19 &53 &18 &6 &4 &0\\
			Class 3 &0 &10 &90 &0 &0 &0 &0 \\
			Class 4 &10 &11 &74 &2 &3 &0 &0\\ 
			Class 5 &0 &5 &85 &3 &7 &0 &0 \\  
			Class 6 &11 &15 &56 &9 &1 &6 &2\\ \hline
		\end{tabular}
	%\end{ruledtabular}
\end{table}

\begin{table}[h]
	\caption{\label{tab:table3}Output of software Mplus for question II.A (in \%)}
 \centering
	%\begin{ruledtabular}
		\renewcommand{\arraystretch}{1.1}
		\begin{tabular}{p{1.1cm} c c c c c c c} \hline
			Class       & a     &b      &c      &d      &e      &f      &x     \\\hline
			Class 1    &0 &10 &62 &0 &23 &5 &0 \\
			Class 2a   &7 &0 &83 &4 &0 &6 &0  \\
			Class 2b    &12 &32 &30 &6 &17 &3 &0 \\
			Class 3     &3 &18 &69 &10 &0 &0 &0 \\
			Class 4     &0 &13 &75 &5 &7 &0 &0 \\ 
			Class 5     &1 &6 &37 &0 &36 &20 &0 \\  
			Class 6     &1 &17 &34 &6 &0 &42 &0 \\ \hline
		\end{tabular}
	%\end{ruledtabular}
\end{table}

\begin{table}[h]
	\caption{\label{tab:table3}Output of software Mplus for question I.B (in \%)}
 \centering
	%\begin{ruledtabular}
		\renewcommand{\arraystretch}{1.1}
		\begin{tabular}{p{1.1cm} c c c c c c c} \hline
			Class       & a     &b      &c      &d      &e      &f      &x     \\\hline
			Class 1 &78 &5 &4 &12 &0 &0 &1 \\
			Class 2a &86 &0 &3 &9 &2 &0 &0 \\
			Class 2b &36 &36 &13 &0 &9 &6 &0 \\
			Class 3 &90 &0 &7 &3 &0 &0 &0 \\
			Class 4 &52 &5 &25 &18 &0 &0 &0 \\
			Class 5 &15 &12 &14 &13 &38 &8 &0 \\
			Class 6 &23 &10 &17 &7 &13 &30 &0 \\ \hline
		\end{tabular}
	%\end{ruledtabular}
\end{table}

\begin{table}[h]
	\caption{\label{tab:table3}Output of software Mplus for question I.C (in \%)}
 \centering
	%\begin{ruledtabular}
		\renewcommand{\arraystretch}{1.1}
		\begin{tabular}{p{1.1cm} c c c c c c c} \hline
			Class       & a     &b      &c      &d      &x        \\\hline
class 1 &89 &0 &8 &3 &0  \\
class 2a &61 &5 &32 &2 &0   \\
class 2b &51 &22 &27 &0 &0   \\ 
class 3 &49 &23 &28 &0 &0   \\
class 4 &38 &0 &62 &0 &0   \\
class 5 &0 &65 &21 &13 &1   \\
class 6 &26 &18 &26 &30 &0   \\ \hline
		\end{tabular}
	%\end{ruledtabular}
\end{table}

\begin{table}[h]
	\caption{\label{tab:table3}Output of software Mplus for question II.C (in \%)}
 \centering
	%\begin{ruledtabular}
		\renewcommand{\arraystretch}{1.1}
		\begin{tabular}{p{1.1cm} c c c c c c c} \hline
			Class       & a     &b      &c      &d      &x         \\\hline
Class 1 &15 &55 &0 &30 &0  \\
Class 2a &51 &0 &38 &11 &0  \\
Class 2b &37 &16 &37 &10 &0  \\
Class 3 &43 &0 &19 &28 &10  \\
Class 4 &18 &23 &59 &0 &0  \\
Class 5 &4 &46 &25 &25 &0  \\
Class 6 &12 &6 &37 &45 &0  \\ \hline
		\end{tabular}
	%\end{ruledtabular}
\end{table}

\vfill

\begin{table}[h]
	\caption{\label{tab:table3}Output of software Mplus for question I.D (in \%)}
 \centering
	%\begin{ruledtabular}
		\renewcommand{\arraystretch}{1.1}
		\begin{tabular}{p{1.1cm} c c c c c c c} \hline
			Class       & a     &b      &c      &d      &e      &f      &x     \\\hline
			Class 1 &57 &31 &0 &0 &1 &11 &0 \\
			Class 2a &64 &30 &0 &1 &5 &0 &0 \\
			Class 2b &32 &31 &9 &3 &0 &25 &0 \\
			Class 3 &0 &17 &26 &34 &9 &14 &0 \\
			Class 4 &30 &59 &0 &8 &1 &2 &0 \\
			Class 5 &19 &45 &1 &0 &14 &21 &0 \\ 
			Class 6 &26 &17 &9 &11 &0 &37 &0  \\ \hline
		\end{tabular}
	%\end{ruledtabular}
\end{table}

\begin{table}[h]
	\caption{\label{tab:table3}Output of software Mplus for question II.D (in \%)}
 \centering
	%\begin{ruledtabular}
		\renewcommand{\arraystretch}{1.1}
		\begin{tabular}{p{1.1cm} c c c c c c c} \hline
			Class       & a     &b      &c      &d      &e      &f      &x     \\\hline
			Class 1 &18 &58 &0 &1 &12 &11 &0 \\
			Class 2a &15 &79 &0 &0 &5 &0 &1 \\
			Class 2b &27 &39 &7 &0 &0 &27 &0 \\
			Class 3 &0 &0 &35 &24 &10 &31 &0 \\
			Class 4 &61 &35 &4 &0 &0 &0 &0 \\
			Class 5 &31 &29 &2 &2 &22 &14 &0 \\
			Class 6 &11 &28 &7 &9 &5 &40 &0  \\ \hline
		\end{tabular}
	%\end{ruledtabular}
\end{table}

\begin{table}[h]
	\caption{\label{tab:table3}Output of software Mplus for question I.E (in \%)}
 \centering
	%\begin{ruledtabular}
		\renewcommand{\arraystretch}{1.1}
		\begin{tabular}{p{1.1cm} c c c c c c c} \hline
			Class       & a     &b      &c      &d      &e      &f      &x     \\\hline
			Class 1 &44 &38 &0 &6 &2 &10 &0 \\
			Class 2a &63 &28 &2 &0 &7 &0 &0 \\
			Class 2b &25 &24 &18 &24 &0 &9 &0 \\
			Class 3 &52 &37 &0 &11 &0 &0 &0 \\
			Class 4 &19 &39 &14 &21 &0 &7 &0 \\
			Class 5 &27 &21 &16 &18 &2 &16 &0 \\
			Class 6 &15 &19 &8 &14 &6 &38 &0 \\ \hline
		\end{tabular}
	%\end{ruledtabular}
\end{table}

\begin{table}[h]
	\caption{\label{tab:table3}Output of software Mplus for question II.E (in \%)}
 \centering
	%\begin{ruledtabular}
		\renewcommand{\arraystretch}{1.1}
		\begin{tabular}{p{1.1cm} c c c c c c c} \hline
			Class       & a     &b      &c      &d      &e      &f      &x     \\\hline
			Class 1 &29 &25 &2 &14 &23 &7 &0 \\
			Class 2a &39 &16 &8 &10 &22 &5 &0 \\
			Class 2b &20 &30 &21 &12 &5 &12 &0 \\
			Class 3 &15 &33 &7 &0 &31 &14 &0 \\
			Class 4 &13 &40 &6 &8 &21 &12 &0 \\
			Class 5 &6 &24 &1 &35 &16 &18 &0 \\
			Class 6 &17 &14 &3 &11 &4 &51 &0 \\ \hline
		\end{tabular}
	%\end{ruledtabular}
\end{table}

\pagebreak
\onecolumn
\section{Appendix C: results per country}

\begin{table}[h]
	\caption{\label{tab:table3}Student scores on the AMoSS test (\textit{N} = 262) per country. If the difference between the pretest and posttest is significant (p $<$ 0.05), this is marked by an asterisk\textsuperscript{*}.}
	\centering
	%\begin{ruledtabular}
	\renewcommand{\arraystretch}{1.3}
	\begin{tabular}{p{2.8cm} p{0.9cm} p{0.9cm} p{0.9cm} p{0.4cm} p{0.9cm} p{0.9cm} p{0.9cm} p{0.4cm} p{0.9cm} p{0.9cm} p{0.9cm}}\hline
		& \multicolumn{3}{c}{Sun questions}    &    & \multicolumn{3}{c}{Star questions} & &\multicolumn{3}{c}{All questions }\\
		&PRE &POST &GAIN & &PRE &POST &GAIN &   &PRE &POST &GAIN\textbf{}
		\\ \hline
		\multicolumn{4}{l}{BELGIUM (\textit{N} = 90)}
		\\
		Mean &54 \%\textsuperscript{*}  &66 \%\textsuperscript{*}  &+12 \% & &30 \%\textsuperscript{*} &47 \%\textsuperscript{*} &+17 \% & &42\%\textsuperscript{*} &56\%\textsuperscript{*} &+14 \%\\
		Median  &67 \%    &67 \% & & &33 \%            &50 \%  & &   &42 \% &58 \% \\
		Standard deviation   &25 \%    &22 \% & & &18 \%    &23 \%  & &   &17 \% &18 \%
		\\ \hline
		\multicolumn{4}{l}{GERMANY (\textit{N} = 30)}
		\\
		Mean &32 \%\textsuperscript{*}  &48 \%\textsuperscript{*} &+16 \% & &21 \%\textsuperscript{*} &34 \%\textsuperscript{*} &+13 \% & &26\%\textsuperscript{*} &41\%\textsuperscript{*} &+15 \%  \\
		Median  &25 \%    &50 \% & & &17 \%            &33 \%  & &   &25 \% &42 \% \\
		Standard deviation   &25 \%    &20 \% & & &18 \%    &22 \%   & &  &18 \% &17 \%
		\\ \hline
		\multicolumn{4}{l}{GREECE (\textit{N} = 46)} \\
		Mean &30 \% \textsuperscript{*}  &41 \% \textsuperscript{*} &+11 \% & &24 \% &32 \% &+8 \% & &27\% \textsuperscript{*} &37\% \textsuperscript{*} &+10 \% \\
		Median  &33 \%    &33 \% & &  &17 \%            &33 \%  & &  &25 \% &33 \% \\
		Standard deviation   &20 \%    &21 \% & & &20 \%    &25 \%  & &   &15 \% &19 \% \\ \hline
		\multicolumn{4}{l}{ITALY (\textit{N} = 96)} \\
		Mean &35 \%\textsuperscript{*}  &53 \%\textsuperscript{*} &+18 \% & &23 \%\textsuperscript{*} &33 \%\textsuperscript{*} &+10 \% & &29\%\textsuperscript{*} &43\%\textsuperscript{*} &+14 \% \\
		Median  &33 \%    &50 \% & & &17 \%            &33 \%   & &  &25 \% &42 \% \\
		Standard deviation   &22 \%    &24 \% & & &15 \%    &19 \%   & &  &15 \% &17 \% \\ \hline
	\end{tabular}
	%\end{ruledtabular}
\end{table}

\begin{figure}[h] % de h staat voor "hier"
	\centering
	\includegraphics[width=17cm]{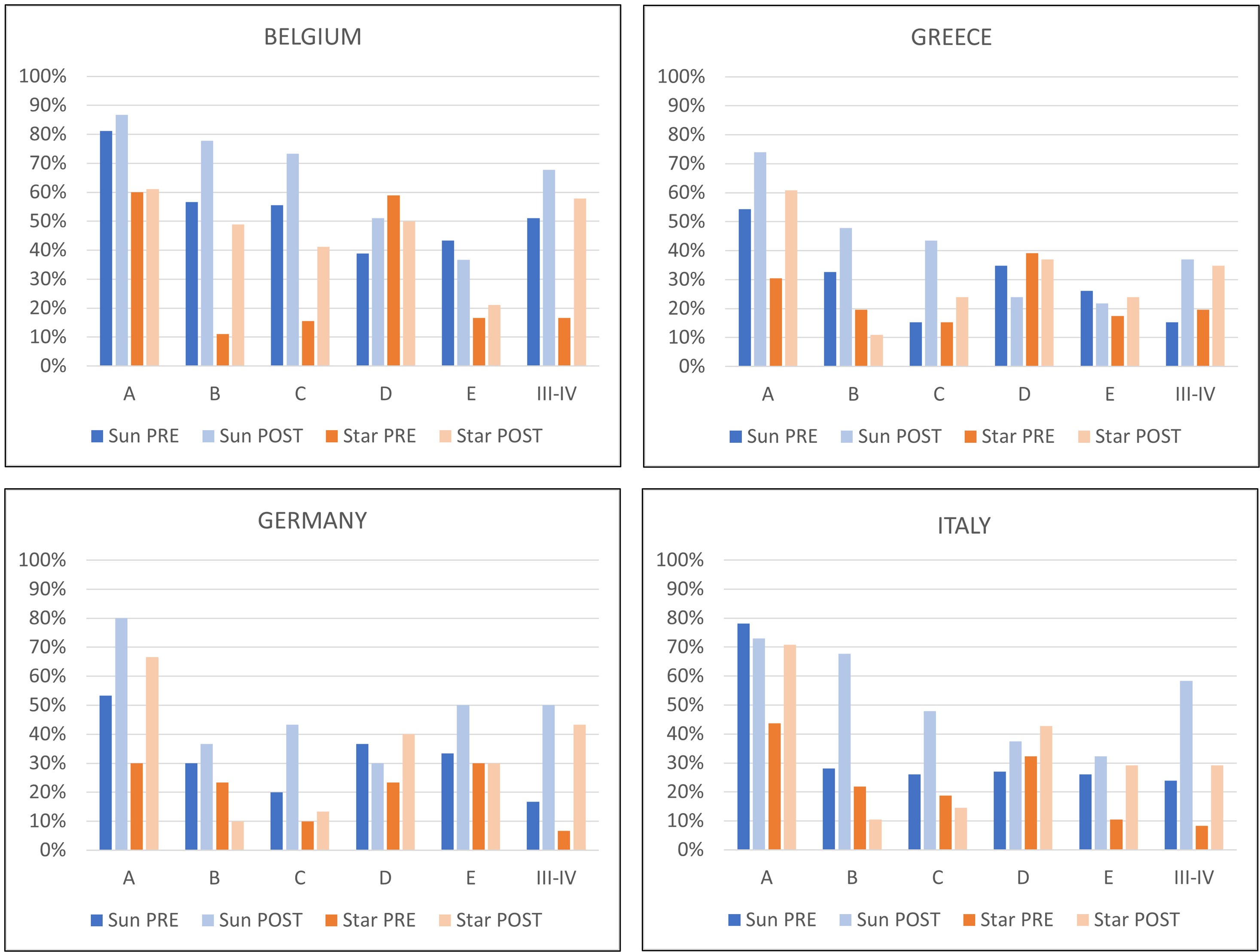}% Here is how to import EPS art
	\caption{\label{fig:epsart} Percentage of students with a correct answer on pretest and posttest, ordered per question.}
\end{figure}

\begin{figure}[h] % de h staat voor "hier"
	\centering
	\includegraphics[width=17cm]{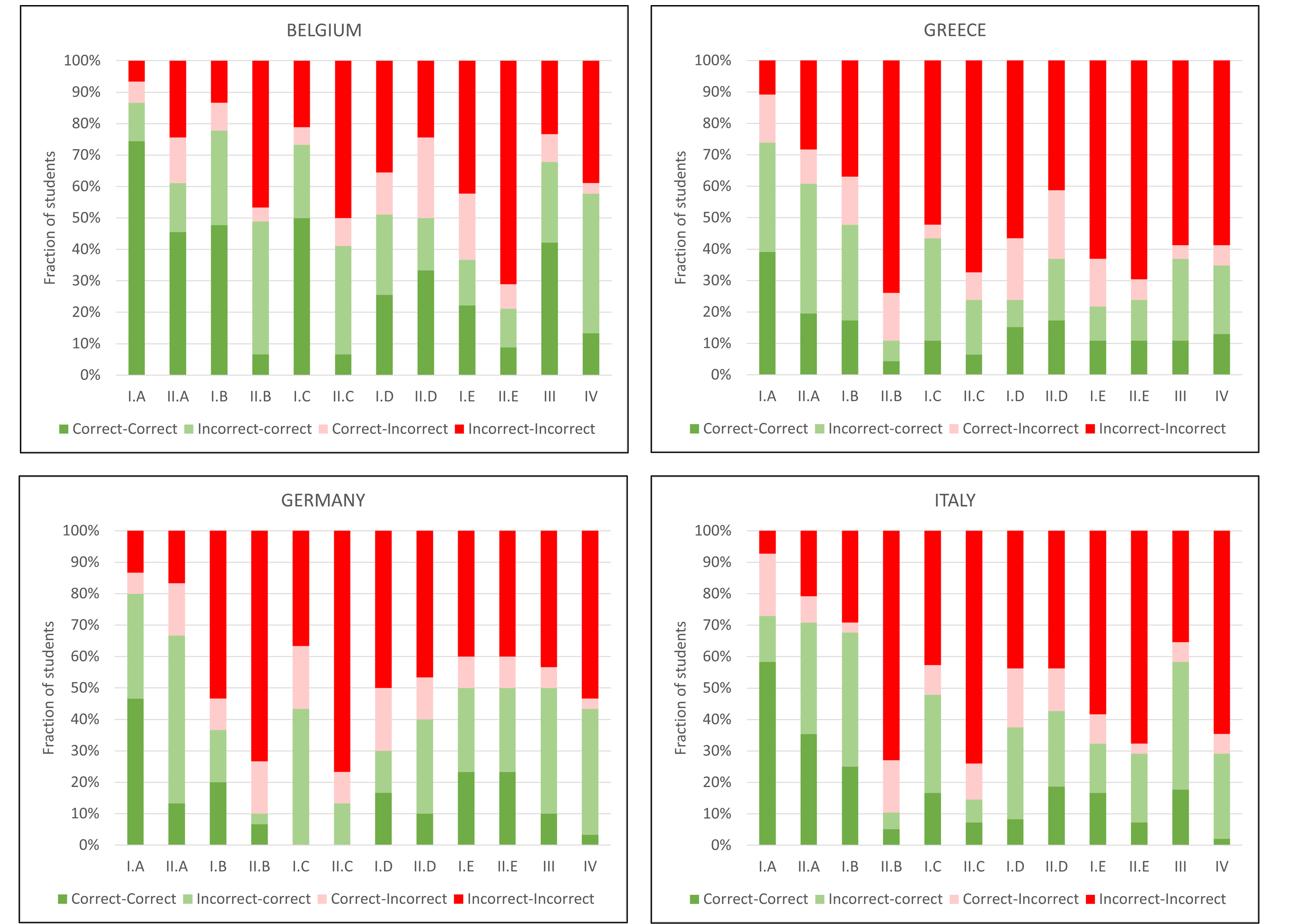}% Here is how to import EPS art
	\caption{\label{fig:epsart} Change of correctness between pretest and posttest, ordered per question.}
\end{figure}

\begin{figure}[h] % de h staat voor "hier"
	\centering
	\includegraphics[width=17cm]{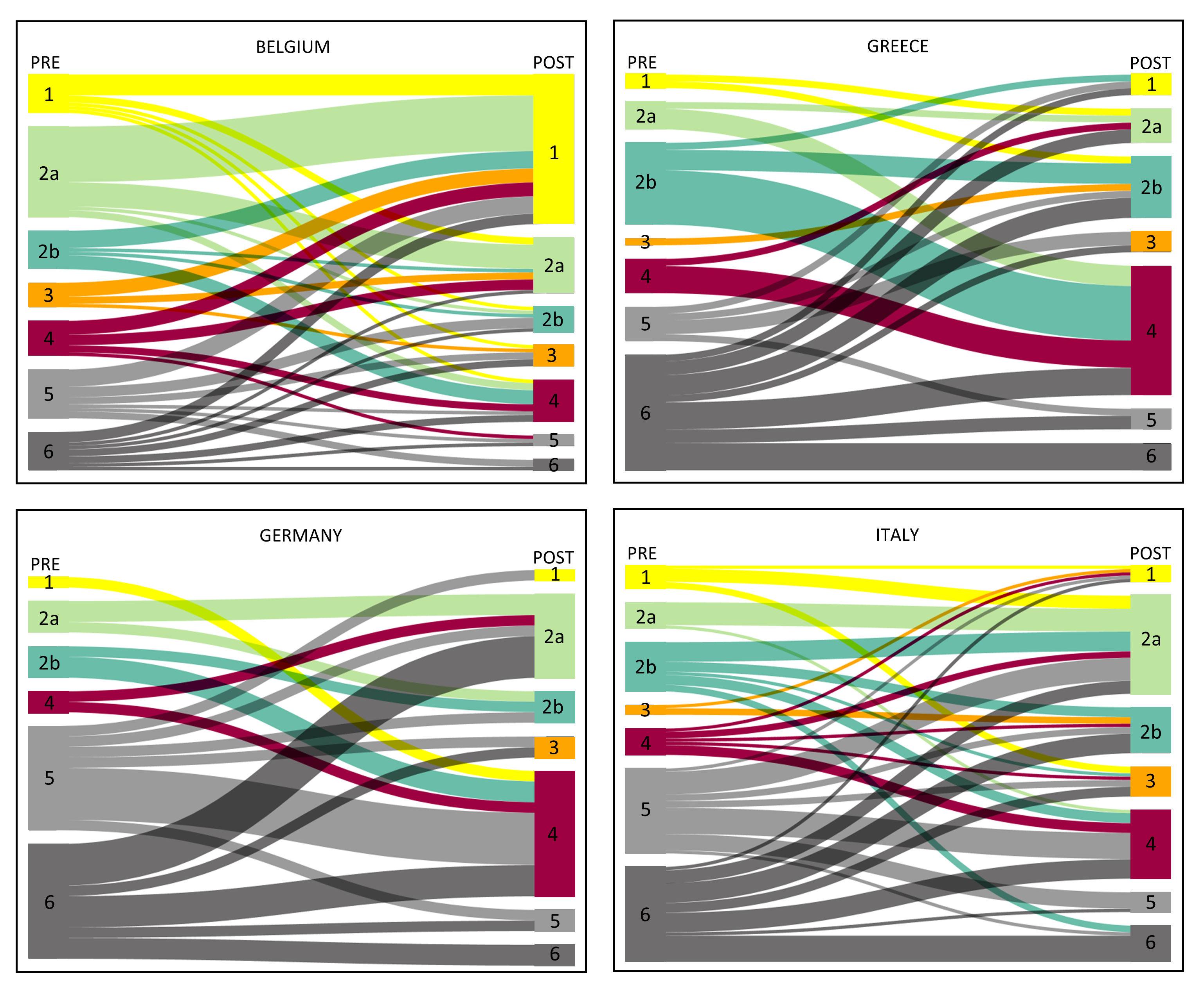}% Here is how to import EPS art
	\caption{\label{fig:epsart} Evolution of class membership between pretest and posttest. The higher the class is positioned in the figure, the better the student scores on the AMoSS test.}
\end{figure}

\end{document}